\documentclass[a4paper,printer]{aa}
\usepackage[varg]{txfonts}

\usepackage[T1]{fontenc}
\usepackage[hidelinks]{hyperref}
\usepackage{graphicx}	% Including figure files
\usepackage{amsmath}	% Advanced maths commands
\usepackage{amssymb}	% Extra maths symbols
\usepackage{microtype,xcolor,tikz}
\usepackage{enumerate}
\usepackage{multicol}

\title{Bars formed in galaxy merging and their classification with deep learning}

\author{M. K. Cavanagh\inst{\ref{inst1}} \and K. Bekki\inst{\ref{inst1}}}

\institute{ICRAR M468, The University of Western Australia, 35 Stirling Hwy, Crawley, Western Australia 6009, Australia\label{inst1}}

\titlerunning{Bar formation with deep learning}
\authorrunning{M. K. Cavanagh \& K. Bekki}

\abstract{Stellar bars are a common morphological feature of spiral 
galaxies.  While it is known that they can form in isolation, or be 
induced tidally, few studies have explored the production of 
stellar bars in galaxy merging.  We look to investigate bar 
formation in galaxy merging using methods from deep learning to 
analyse our N-body simulations.}
{The primary aim is to determine the constraints on the mass ratio 
and orientations of merging galaxies that are most conducive to bar 
formation.  We further aim to explore whether it is possible to 
classify simulated barred spiral galaxies based on the mechanism of 
their formation.  We test the feasibility of this new 
classification schema with simulated galaxies.}
{Using a set of 29,400 images obtained from our 
simulations, we first trained a convolutional neural network to 
distinguish between barred and non-barred galaxies.  We then tested 
the network on simulations with different mass ratios and spin 
angles. We adapted the core neural network architecture for use 
with our additional aims.}
{We find that a strong inverse relationship between mass ratio and 
the number of bars produced.  We also identify two distinct phases in 
the bar formation process; (1) the initial, tidally induced 
formation pre-merger, and (2) the destruction and/or regeneration 
of the during and after the merger.}
{Mergers with low mass ratios and closely-aligned orientations are 
considerably more conducive to bar formation compared to equal-mass 
mergers. We demonstrate the flexibility of our deep learning 
approach by showing it is feasible to classify bars based on their 
formation mechanism.}


\begin{document}

\maketitle

\section{Introduction}
Central bar structures are a common feature of most spiral 
galaxies. Around 30\% of all spiral galaxies are understood to be 
strongly barred \citep{sellwood-1993}, while overall more than 50\% 
of spiral galaxies typically contain features indicative of a 
central bar structure \citep{eskridge-1999}. Bars are found to be 
even more prevalent when observed in infrared 
\citep{eskridge-2000}. Previous studies into the significance and 
physical properties of central bar structures have found that bars 
play an important role in both the long and short-term evolution of 
spiral galaxies. It is known that stellar bars have a profound 
impact on the dynamics and evolution of spiral galaxies 
\citep{little-1991, sellwood-1993, athana-2005, vera-2016},
the subsequent evolution of the interstellar medium 
\citep{mayer-2006, fanali-2015}.  Bars also drive interactions 
between the stellar disk and dark matter halo
\citep{athana-2003, dalcanton-2004, sellwood-2014},
redistribute angular momentum within the disk
\citep{athana-2005, aguerri-2009} and accelerate the
formation of spiral structure
\citep{lynden-bell-1972, pfenniger-1991}, central bulges
\citep{kormendy-2004, gadotti-2011} and other morphological 
features \citep{rautiainen-2002, regan-2003}.

In this work we consider three models of bar formation; the self-
gravitating isolated model (also known as the spontaneous model), 
the tidal interaction model, and the galaxy merger model using N-
body simulations based on the previous work of 
\citet{bekki-2013, bekki-2019}. The physical conditions behind the formation of isolated bars and tidally-induced bars have already been well investigated
\citep{hohl-1971, little-1991, noguchi-1987, barnes-1992, 
miwa-1998, berentzen-2004}, however there has been limited to no 
research on the formation of bars in galaxy merging.  Galaxy merging is an important factor in the evolution of galaxies \citep{conselice-2014}, particularly as it affects star formation \citep{pearson-2019} and the growth of galaxies \citep{cattaneo-2011}.

The primary purpose of this paper is to investigate the physical 
parameters behind the formation of stellar bars in galaxy merging.  
Studies such as \citet{peirani-2009, moetazedian-2017} have shown 
that merging galaxies can induce bars prior to the collision, but 
we look to investigate the circumstances under which bars can 
emerge from the aftermath of a merger unscathed.
We investigate the effect of the mass-ratio $m_2$ and spin angles $
\theta_1, \theta_2$ of two merging galaxies on the incidence of bar 
formation (these are formally defined in \S 3.3).  Aided by deep 
learning, we attempt to determine the physical parameters most 
conducive to bar formation. We propose a method to use 
convolutional neural networks (CNNs) to automatically classify the 
images from the results of our N-body galaxy simulations as either 
barred or non-barred. Such a method allows us to rapidly classify 
data and enables data otherwise considered intractable to be 
analysed in a more practical time frame. 
Apart from the initial work needed to prepare and label  
the training data, the neural network is able to quickly test our 
simulation outputs.

In addition to our primary aim, we also look to investigate the 
ability of a CNN to discriminate
between multiple mechanisms of bar formation: i.e, given an image 
of a galaxy, determine whether it was formed (a) in isolation, (b) 
due to the tidal interaction or (c) during a galaxy merger. 
Since CNNs work by identifying features in data, this feasibility 
study is intended to determine whether a barred 
galaxy's current morphology is indicative of the means by which it 
was formed.  If so, this has the potential to further accelerate 
large-scale surveys into the prevalence and nature of bars in the 
Universe.

The structure of the paper is as follows.
We provide a brief background behind the 
three formation mechanisms and our neural network implementation in \S 2.
We explain our simulation models and orbital parameters, as well as 
describe the process of our CNN-based analysis in \S 3.
We present the results of our analysis in in \S 4.
In \S 5 we discuss our results, focusing on the effect of mass 
ratio and spin angles of the overall bar formation process
In \S 6, we discuss our secondary aim of using a CNN to distinguish 
between multiple bar formation mechanisms, along with an evaluation 
of our CNN approach.
Lastly, we summarize our key conclusions in \S 7.

\section{Bar formation mechanisms and deep learning}

\subsection{Bar formation mechanisms}

In this work we consider three mechanisms of bar formation: the 
isolated or self-gravitating model, the tidal interaction model and 
the galaxy merger model.

\subsubsection{Isolated model}

It is well known that large-scale self-gravitating disks are prone 
to instability unless certain conditions for stability are met 
\citep{toomre-1964}. The first two-dimensional numerical 
simulation to investigate disk instabilities was conducted by 
\citet{hohl-1971}, who found that slow-growing large-scale non-
axisymmetric disturbances eventually led to the formation of a 
central bar structure. Later simulations conducted in three 
dimensions were able to observe bars forming from initially 
axisymmetric disks suggesting that the global instability is an 
inherent property of the disk, regardless of whether or not the 
initial particle distributions are axisymmetric. The 3D simulation 
by \citet{pfenniger-1984} found significant vertical instability 
despite the flat disk. Interestingly, this vertical instability 
can lead to dynamical buckling and the formation of boxy peanut-
shaped bars \citep{raha-1991}. \citet{sparke-1987} notes the 
difficulty for a completely isolated bar to survive indefinitely. 
Indeed, simulations show that such isolated bars can be destroyed 
due to large-scale accretion \citep{friedli-1993}. It has long 
been established that bars can form due to orbital resonances 
associated with global instabilities 
\citep{athana-1983,contopoulos-1989,pfenniger-1991}. Several 
mechanisms have been proposed to explain the how these 
instabilities result in bar formation, such as the swing 
amplification of propagating density waves \citep{goldreich-1965, 
julian-1966} to the Lynden-Bell mechanism of resonant orbits 
\citep{lynden-bell-1979}.

\subsubsection{Tidal model}

The isolated model is a case of internally-induced bar formation. 
However, bars are more frequently observed in dense environments 
such as clusters \citep{elmegreen-1990} with characteristics that 
differ from isolated bars, such as a slower pattern speed
and differences in the location of inner Lindblad resonances 
\citep{miwa-1998}. It has long been recognised that tidal 
interactions between galaxies can influence morphology and bar
formation \citep{toomre-1972,athana-1999,barnes-1992}.
The pioneering simulation of 
\citep{noguchi-1987} investigating the effects of tidal deformation 
of galaxy disks and showed that non-axisymmetric instabilities can 
be induced throughout the disk, leading to the formation of a 
stellar bar. Importantly, the strength of this bar was found to 
depend on many parameters, such as the mass ratio between the 
perturbed and peturbing galaxies, as well as bulge and halo ratios 
\citep{noguchi-1987}. Tidal interactions are sufficient to support 
the evolution of the bar, with the capability to regenerate the bar 
should it weaken due to its inherent instability 
\citep{berentzen-2004}. This can be simulated in the case of low-
mass satellite galaxies interacting with their host, where the 
properties of the host's tidal bar are found to be independent of 
the number of tidal interactions \citep{moetazedian-2017}.

\subsubsection{Merger model}

Galaxing merging is a form of direct galaxy-galaxy interaction 
where two galaxies come into contact. However, the large timescale 
over which such mergers take place necessitate the use of numerical 
simulations in order to investigate morphological changes and the 
possibility of bar formation. Previous research has shown that 
significant changes to morphology and star formation can be 
triggered due to galaxy merging \citep{hernquist-1995}.
Galaxy merging is a particularly important driver of star formation 
\citep{lin-2007, bridge-2007, kaviraj-2009, kaviraj-2015} and also plays a 
key role in the morphological evolution of galaxies 
\citep{conselice-2014}.
The traditional view is that many galaxy mergers, particularly 
major mergers where the masses of the colliding galaxies are 
approximately equal, destroy the overall galaxy disk with the 
remnants eventually combining to form early-type elliptical 
galaxies \citep{toomre-1972,negroponte-1983,barnes-1996}. Despite 
this, more recent simulations have shown that spiral galaxies can 
emerge from galaxy mergers \citep{springel-2005} as well as the 
formation of stellar bars \citep{di-matteo-2010}. Our work aims to 
guide future exploration of the latter, for although numerical 
simulations have demonstrated that bars can be formed in galaxy 
merging \citep{peirani-2009, lotz-2010}, the processes behind 
bar formation in mergers are not well understood. It is important to stress that \citet{peirani-2009} used a merger to model a particular galaxy's stellar bar, while in this work we focus on the mechanism of bar formation through simulating many hundreds of mergers in order to systematically test the effects of changing the mass ratio $m_2$ and spin angles $\theta_1, \theta_2$.

\subsection{Convolutional neural networks}

In the literature, a variety of machine learning methods have been 
applied in the field of astronomy, ranging from artificial neural 
networks with the aim to classify galaxy morphology
\citep{naim-1995, calleja-2004, dieleman-2015, consolandi-2016, 
abraham-2018, diaz-2019}, to analysing large-scale simulations \citep{huertascompany-2019} and to general-purpose training programs \citep{vasconcellos-2011, graff-2014}. 
These applications of deep learning for morphological 
classification are not restricted to the optical, with 
\citet{wu-2018} and \citet{lukic-2019} investigating morphological 
classification of radio sources.

In this work, we utilise convolutional neural networks to 
determine the presence of a bar in an image of a galaxy, and later 
classify a bar according to the mechanism with which it was formed. A CNN \citep{lecun-1998} is a type of neural network 
that is especially useful for image 
analysis due to its use of feature maps \citep{lecun-2015}. These 
feature maps excel at finding high-level features of an 
image, which are key to the ability to discriminate one image from 
another \citep{haykins-2009}.

\begin{figure*}
\centering
\includegraphics[scale=0.5]{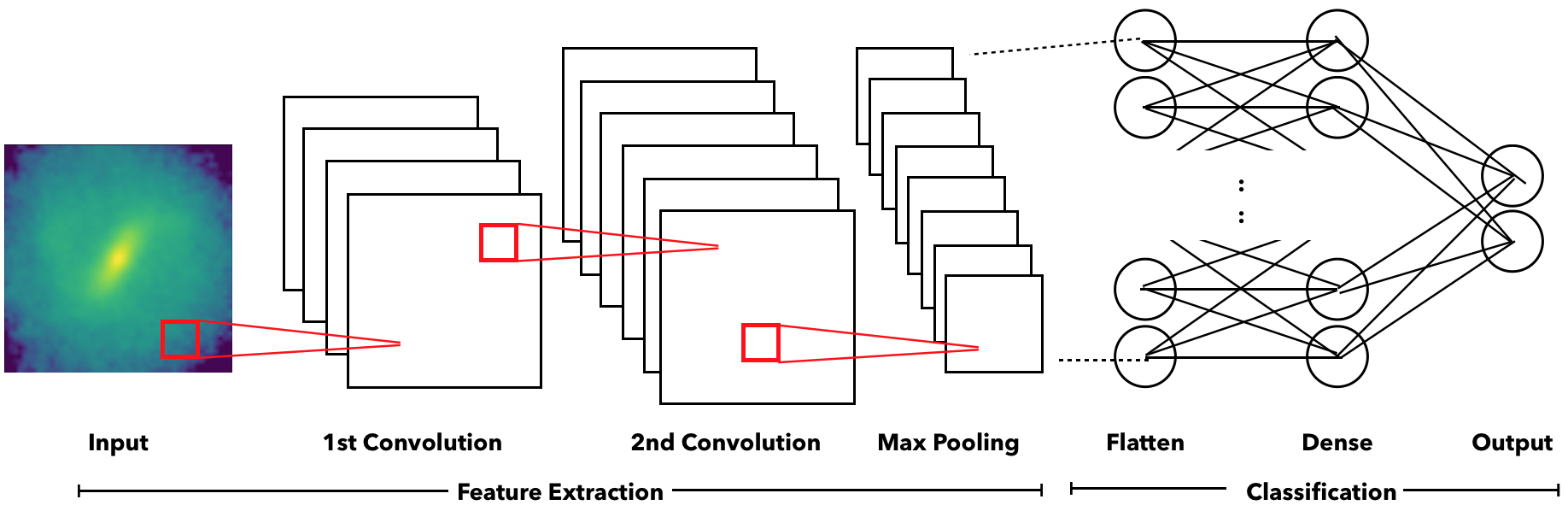}
\caption{Schematic overview of the general architecture of a 
convolutional neural network. The input image is mapped into 
various feature maps via convolutions, followed by a pooling layer 
that performs local averaging and subsampling. The final layer is 
a list of outputs that, in our case, correspond to whether a bar is 
present or not.}
\label{fig:cnn-schematic}
\end{figure*}

A general schematic showing how a CNN processes an input image is shown in Figure \ref{fig:cnn-schematic}. There exist many different 
CNN architectures in the literature \citep{lecun-2015} that differ 
in the structure, type and quantity of component layers, as well as 
utilising different weighting and activation functions. However, memory and (computational) time limitations ultimately result in some degree of compromise between complexity and overall performance.

Our CNN architecture is based on \citep{bekki-2019}, with the goal 
of inputting a 50x50 density map and outputting a label to indicate 
whether or not a bar was detected. We used Keras for the CNN 
construction \citep{chollet-2015}. There are 2 convolution layers 
(\texttt{Conv2D}), a max-pooling layer (\texttt{MaxPooling2D}), two 
dropout layers (\texttt{Dropout}), a flatten layer 
(\texttt{Flatten}) and finally two dense layers (\texttt{Dense}) 
that are fully-connected (including the output layer). The role of the convolution layers is to extract the high-
level features of the preceeding layer. This is done via the use 
of a kernel filter that performs a matrix operation, progressively 
scanning through the input layer. In our network, each 
\texttt{Conv2D} layer uses a 3x3 kernel. Our 50x50 
input image is mapped into 32 separate convolved feature maps 
via the first \texttt{Conv2D} layer. These maps are then further 
convolved into 64 separate feature maps via the second 
\texttt{Conv2D} layer.  Each of these convolutional layers can be 
thought of as a layer of abstraction, enabling higher-level 
features to be extracted.

The next major part of the architecture is the 
\texttt{MaxPooling2D} layer. In general, a pooling layer is 
designed to average out the convolved feature maps through 
downsampling. Instead of averaging, \texttt{MaxPooling2D} returns 
local maxima. This achieves two major goals: firstly, it decreases 
the computational complexity, while secondly the max-pooling 
retains the dominant features from the convolved feature map. The 
\texttt{Flatten} layer is used to vectorise 
the feature maps, while the \texttt{Dropout} layers are used to 
avoid over-fitting by ignoring a fraction of the preceding nodes. The final
fully-connected \texttt{Dense} layer (before the output layer) thus 
combines the flattened nodes.
From here on, the architecture is similar to that of a basic neural 
network; the weights and biases of the \texttt{Dense} layer nodes 
can be tweaked in order to associate different high-level features 
with the desired output category.
For our CNN, we used the ReLu activation function for 
the convolution and dense layers, whilst the final output layer 
used a softmax (normalised exponential) activation with the categorical cross-entropy loss function.
For training, we used the ADADELTA 
adaptive learning rate method, developed by \citet{zeiler-2012}.
This core architecture was used for both parts of our
investigation, the only crucial difference being that the CNN used
to test multiple bar formation mechanisms had an extended output
layer in order to account for the new labelling scheme.

\subsection{Differences with existing bar detection techniques}

There are several existing methods in which to 
both detect bars and analyse their physical characteristics; 
ellipse fitting of the galaxy isophotes, Fourier analysis of the 
(azimuthal) luminosity profile and a decomposition of the surface 
brightness distribution \citep{prieto-2001}. These methods involve 
significant pre and post-processing, and often require access to 
additional data such as spectroscopic flux measurements and 
velocity dispersion maps.  Hence these methods cannot readily 
determine the presence of a bar from mere images alone.  Another 
method involves directly analysing the image of a galaxy using 
Fourier techniques.  These methods are well-known to detect the 
presence of stellar bars and spiral arms, as well as characterise 
their shape and strength \citep{garcia-gomez-1991, aguerri-1998, 
barbera-2004, garcia-gomez-2017}.

CNNs are similar to Fourier transforms in this regard,
as their core operation 
revolves around the detection of high-level features.
Morphological classification is thus a natural application.  
Another difference with our CNN approach is that it avoids much of 
the extra image processing needed in traditional methods more 
suited to observational data (see \citep{prieto-2001}).  
Furthermore, the CNN can be trained on images with high 
inclination, whereas methods similar to \citet{garcia-gomez-2017} 
must perform a de-projection.

The bottleneck for CNNs lies with the training. Once trained, CNNs 
can quickly analyse images and output a label corresponding to what 
it has detected the image to be. Existing methods are tailored for 
real-world data, whereas our CNN is specifically designed to work 
with the output of our galaxy simulations.  It is computationally 
expensive to accurately model luminosity profiles.  Our CNN works 
with the output of dynamical N-body simulations, allowing it to be 
trained and tested on a larger dataset than we would otherwise have 
been able to compile. Our CNN-based approach to bar detection is 
merely tackling the problem from a different angle (i.e feature 
detection with CNNs) where all we are given is the image of the 
galaxy.  It is not designed to supplant nor contest any existing 
method; rather it is a tool well suited to our aim of analysing 
many galaxy simulations in order to constrain parameters based on 
how many bars are detected.

\section{Simulation and implementation}

\subsection{Structure and kinematics of the stellar disk}

We consider merging of two spiral galaxies with various bulge-to-
disk-ratios and baryonic mass fractions
in order to investigate how stellar bars can be induced during 
galaxy merging. The total masses of dark matter halo, stellar disk, 
gas disk, and bulge of a disk galaxy are denoted as $M_{\rm h}$, $M_{\rm s}$, $M_{\rm g}$,
and $M_{\rm b}$, respectively.
In these preliminary works, we only show the results for models with no gas ($f_{\rm g}=M_{\rm g}/M_{\rm s}=0$)
The bulge-to-disk-ratio is defined as $M_{\rm b}/M_{\rm s}$ and 
represented by the parameter $f_{\rm b}$. 
The key parameters in the present study are 
$M_{\rm h}$, $f_{\rm bary}$, and $f_{\rm b}$.
We adopt the density distribution of the NFW halo \citet{nfw}, as 
suggested from CDM simulations, to describe the initial density 
profile of the dark matter halo in a disk galaxy:
\begin{equation}
{\rho}(r)=\frac{\rho_{0}}{(r/r_{\rm s})(1+r/r_{\rm s})^2},
\end{equation}
where $r$, $\rho_{0}$, and $r_{\rm s}$ are
the spherical radius, the characteristic density of the halo, and 
the scale length of the halo respectively.
The $c$-parameter ($c=r_{\rm vir}/r_{\rm s}$, where $r_{\rm vir}$ 
is the virial radius of a dark matter halo)
along with $r_{\rm vir}$ are chosen appropriately for a given dark 
halo mass($M_{\rm dm}$) by using the $c-M_{\rm h}$ relation for 
$z=0$ predicted by recent cosmological simulations, e.g 
\citet{neto-2007}.

The bulge of the disk galaxy has a size $R_{\rm b}$
and a scale-length $R_{\rm 0, b}$, and is represented by the 
Hernquist density profile.
The bulge is assumed to have an isotropic velocity dispersion,
with radial velocity dispersion given by the Jeans equation
for a spherical system \citep{binney-2008}.
The bulge-to-disk ratio ($f_{\rm b}=M_{\rm b}/M_{\rm d}$) of the 
disk galaxy is a free parameter ranging from 0 (pure disk galaxy) 
to 1.
Our `MW-type' models are those
with $f_{\rm b}=0.17$ and $R_{\rm b}=0.2R_{\rm s}$,
where $R_{\rm s}$ is the stellar disk size of a galaxy.
We adopt the mass-size scaling relation of 
$R_{\rm b} = C_{\rm b} M_{\rm b}^{0.5}$
for bulges such that we can determine $R_{\rm b}$ for a given 
$M_{\rm b}$.
The value of $C_{\rm b}$ is determined in order for $R_{\rm b}$ to 
be 3.5 kpc for $M_{\rm b}=10^{10} {\rm M}_{\odot}$ (this 
corresponds to the mass and size of the MW's bulge).

The radial ($R$) and vertical ($Z$) density profiles of the stellar 
disk are assumed to be proportional to $\exp (-R/R_{0}) $ with 
scale length $R_{0} = 0.2R_{\rm s}$ and to ${\rm sech}^2 (Z/Z_{0})$ 
with scale length $Z_{0} = 0.04R_{\rm s}$, respectively.
In the present model for the MW-type, the exponential disk
has $R_{\rm s}=17.5$ kpc.
In addition to the rotational velocity caused by the gravitational 
field of disk, bulge, and dark halo components, the initial radial 
and azimuthal velocity dispersions are assigned to the disc 
component according to the epicyclic theory \citep{toomre-1964} 
with Toomre's parameter set to $Q$ = 1.5.
The vertical velocity dispersion at a given radius is set to be 0.5
times as large as the radial velocity dispersion at that point.

The total numbers of particles in a fiducial model with 
$f_{\\rm b}=0.167$ is 216700, though for other models it depends on 
$f_{\rm b}$. The mass resolution for each of the models in the 
present study is $1.2 \times 10^6 {\rm M}_{\odot}$.
The gravitational softening length for each component is determined 
by the number of particles used for each component, as well as by 
the size of the distribution
(e.g., $R_{\rm s}$ and $r_{\rm vir}$). It is set to be 320 pc, 
which is much finer than $1-2$ kpc spatial resolutions used for the 
image analysis in the present study.
These spatial and mass resolutions are not particularly high; this 
is predominantly because
we have to run a large number of models within a limited amount of 
GPU computing time allocated for this project.
We believe that the adopted resolutions and subsequent galaxy 
images are sufficient for use with the CNNs.  A summary of our 
simulation parameters and features is provided in Table 
\ref{tab:mwmodel}.

\begin{table}
\centering
\caption{Description of the basic parameter values
for the MW-type disk model.
}
\label{tab:mwmodel}
\begin{minipage}{80mm}
\begin{tabular}{ll}
{Physical properties}
& {Parameter values}\\
\hline
Total halo mass (galaxy)
& $M_{\rm dm}=1.0 \times 10^{12} {\rm M}_{\odot}$ \\
DM structure (galaxy)
& NFW profile \\
Galaxy virial radius (galaxy)
& $R_{\rm vir}=245$ kpc \\
$c$ parameter of galaxy halo
& $c=10$ \\
Stellar disk mass & $M_{\rm s}=6.0 \times 10^{10} {\rm M}_{\odot}$   \\
Stellar disk size & $R_{\rm s}=17.5$ kpc \\
Disk scale length & $R_{0}=3.5$ kpc \\
Gas fraction in a disk & $f_{\rm g}=0$ \\
Bulge mass &  $M_{\rm b}=10^{10} {\rm M}_{\odot}$ \\
Bulge size & $R_{\rm b}=3.5$ kpc \\
Mass resolution & $1.2 \times 10^6 {\rm M}_{\odot}$ \\
Size resolution & 320 pc \\
Spatial resolution for image analysis & 1-2 kpc \\
Star formation & Not included \\
Chemical evolution & Not included \\
Dust evolution & Not included \\
\end{tabular}
\end{minipage}
\end{table}

\subsection{Orbital configurations for galaxy merging}

In all of the merger simulations with different mass-ratios of two 
disks ($m_2$), the orbit of the two disks is set to be initially in 
the xy plane.  The initial distance between the center of mass of 
the two disks, the pericenter distance, 
and the circular velocity factor ($f_{\rm v}$) for the companion 
galaxy are set to be $8R_{\rm s}$, $2R_{\rm s}$, and 0.5 
respectively for all models.
For this study, we have set the velocity of the two disk such that 
all orbits are prograde and initially confined to the xy plane.
We define $f_{\rm v}$ as the ratio of the total 3D velocity
of the companion galaxy (with respect to the primary galaxy)
to the circular velocity at the distance
of the companion from the center of the primary galaxy. The 
azimuthal angle $\phi$ is measured from the x-axis to the 
projection of the angular momentum vector of
the disk onto the x-y plane.  This is set to be 0 for both disk 
galaxies, as it is not an important parameter in the present study.
The time when the progenitor disks merge completely and reach the 
dynamical equilibrium is typically less than 12.0 in our units for 
most of models.

\subsection{Definition of mass ratio and spin angles}

We define the mass ratio $m_2$ as the ratio of the mass of the 
companion galaxy to that of the parent.  We refer to a minor merger 
as a merger with $m_2 \approx 0.1$, and a major merger as that with 
$m_2 \approx 1$.  We also refer to intermediate mass mergers as 
those with $m_2$ between 0.1 and 1. The spin of each galaxy is 
specified by two angles, $\theta_i$, where the subscript i is used 
to identify each galaxy ($\theta_1$  for the primary galaxy, $
\theta_2$ for the companion). The spin angles $\theta_i$ are 
defined as the angle between the z-axis and the vector of the 
angular momentum of the disk.  In our present study, we vary these 
angles between $-180$ and $180$ degrees.

\subsection{Simulation output and training data}

\begin{figure*}
\centering
\includegraphics[scale=0.52]{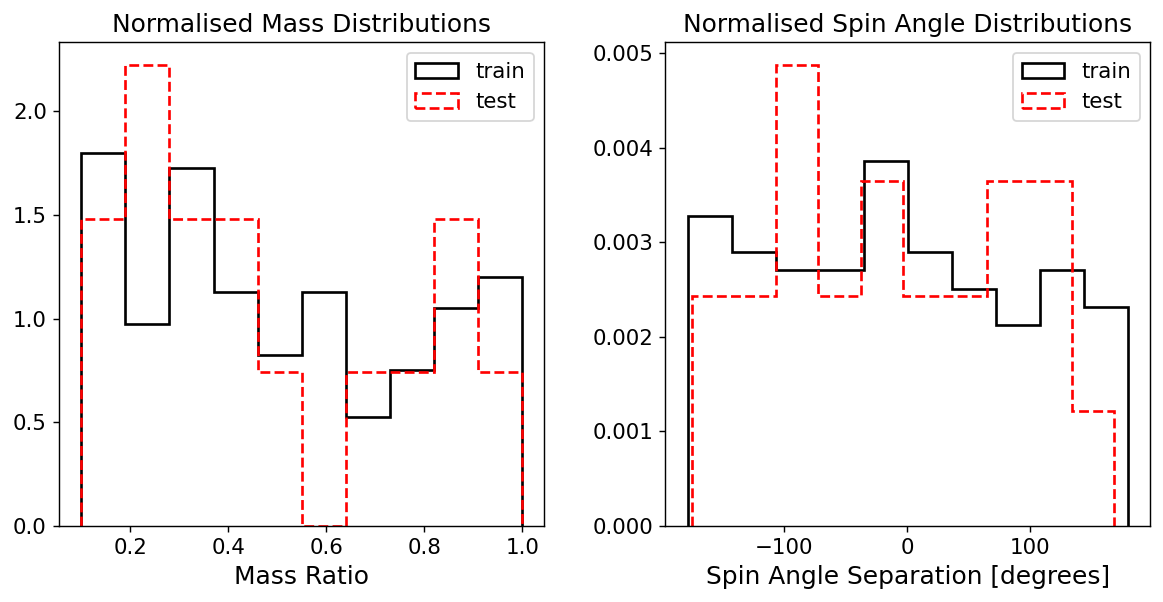}
\caption{Normalised histograms showing the distributions of mass 
ratios (left) and $\theta_1 - \theta_2$ (right) for the training 
and testing sets.}
\label{fig:trset}
\end{figure*}

The raw density output from our simulations comprised a list of 
local density values ordered by pixel for each of the 200 different
viewing angle orientations, with separate files for each model
and time-step. We derive 2D density maps of our simulated galaxies
for $R \le R_{\rm s}$.
We first divide the stellar disk ($R \le R_{\rm s}$)
of the galaxy into a $50 \times 50$ grid mesh and estimate the mean 
stellar mass density at each mesh point as follows:
\begin{equation}
\Sigma_{i,j,0} = \frac{ 1 }{ {\Delta R_{i,j}}^2 } \sum_{k=1}^{N_{\it i,j}}
m_{k},
\end{equation}
where $\Delta R_{i,j}$,
$N_{i,j}$, and $m_{k}$
are the mesh size at the mesh point ($i$, $j$),
the total number of stellar particles in the mesh,
and the mass of a stellar particle, respectively. These values 
were normalised to obtain the normalised density maps. The mesh 
size is set to $0.04 R_{\rm s}$, which roughly corresponds to 0.7 
kpc for an MW-type disk galaxy.

To create the training data for our CNN, the raw
density data was stitched together to create 50x50 images 
(corresponding to the normalised density maps).  Thus the input 
layer to our neural network consisted of a linear
array of 2500 pixel values corresponding to the 50x50 density
maps. This is how the individual images were tested with the
neural network. For each individual galaxy model,
we obtained a total of 200 images that differed in 
orientation and inclination. This is to ensure that the training 
set includes images of bars with different projections and 
orientations, and to mirror real life data.

These density maps, totalling 29,400 total images, were 
combined to create the training set for the CNN. Each of these 
training images were manually assigned a label (corresponding to a 
bar or no-bar) through visual inspection in order to create the input-output pairs with which to train the CNN.
The visual classification was conducted by both authors, restricted to only the face-on images for each model.
On average, 1 in 20 images were classified different by each observer.  These were overwhelmingly galaxies with oblate bulges, where the distinction between a bar and a bulge is not so objectively clear.  As a result, we estimate that the overall labelling of the training set is 95\% accurate (i.e there is an overall uncertainty of around 5\%).
A key aspect of any training set is that should 
be a roughly equal number of samples for each of the classification 
categories, otherwise there is a bias towards the category with the 
greater number of samples.  In our case, that means ensuring the 
number of barred and non-barred galaxy samples are about the same.  
Many of our simulations were thus manually checked to see whether 
or not a bar had been formed. The training set was split into separate training and validation 
sets such that 80\% of the samples were used to train, with the 
remaining 20\% used for validation.
The validation data was randomly partitioned from the main sample set.  Figure \ref{fig:trset} shows the distribution of mass ratio and angular separation ($\theta_1-\theta_2$) for the training and test sets.  Ideally these should be uniform, although there is a slight bias towards samples with lower mass ratios samples.

We define the prevalence of bar formation in a
given model as the fraction of the 200 images that returned a label
of ``bar'' when tested with the CNN. We performed these tests for
density outputs with different values of mass-ratio $m_2$ and spin-
angles $\theta_1$ and $\theta_2$, for different galaxy models with
different bulge-to-disk ratios and dark matter contents. A full
description of the exact parameters used for each model is
provided in Table \ref{tab:models}.

\begin{table*}
\centering
\caption{Dark matter, stellar, bulge components and bulge-to-disk 
ratios for each of our six models}
\label{tab:models}
\begin{tabular}{|c|c|c|c|c|c|c|c|}
\hline 
Model & Dark Matter & Stellar Mass & Bulge Mass & Bulge-to-disk \\ 
\ & $10^{12} M_\odot$ & $10^{10} M_\odot$ & $10^{10} M_\odot$ & ratio \\
\hline
m1 & 1 & 6 & 1 & 0.167 \\ 
\hline 
m2 & 1 & 3 & 0.5 & 0.167 \\ 
\hline 
m3 & 1 & 1.8 & 0.3 & 0.167 \\ 
\hline 
m4 & 1 & 6 & 3 & 0.5 \\ 
\hline 
m5 & 1 & 6 & 6 & 1 \\ 
\hline 
m6 & 1 & 6 & 0 & 0 \\ 
\hline 
\end{tabular} 
\end{table*}

\subsection{Training the network}

\begin{figure*}
\centering
\includegraphics[scale=0.53]{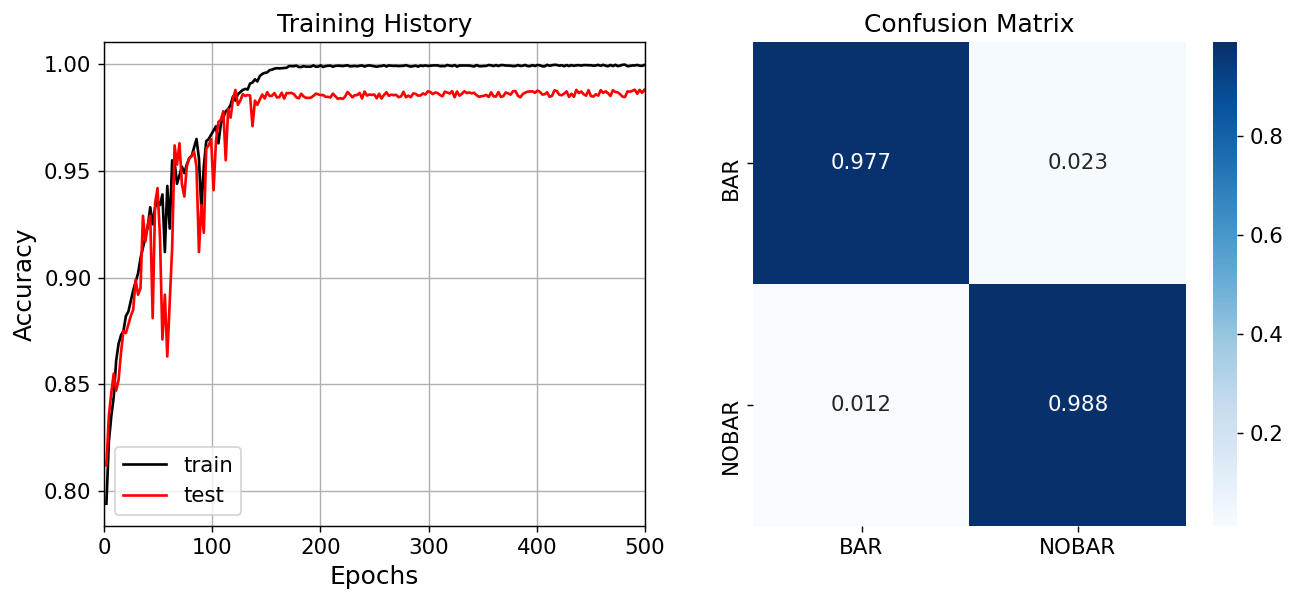}
\caption{Plot of the training and testing (validation) accuracies against training time (left), as well as the confusion matrix (right) for the main CNN}
\label{fig:histconf}
\end{figure*}

The CNN was coded using Keras \citep{chollet-2015}, a high-
level deep learning application programming interface (API), 
running on TensorFlow, a machine learning framework. 
Training was conducted on the ICRAR's Pleiades cluster, making use 
of a Nvidia GTX1080. 
Here all 29,400 images were used to train the CNN, with
training conducted in stages up to a combined total 1000 epochs.  The training and validation accuracies, along with a confusion matrix, are presented in Figure \ref{fig:histconf}.

The validation accuracy converged to between 98\% and 99\% within 
around 200 epochs, after which there was no further increase. Given that this is well above what would be expected due to the 5\% uncertainty inherent in the training set, it is likely that there is some degree of overfitting.
The batch-size was set to 200 images, given that this corresponds to a single galaxy model. Modifying the batch size did not appreciably impact the final accuracy. Training for our secondary, multiple-bar classification CNN was also conducted on Pleiades, however this used a much smaller subset of 4800 images.

To achieve our primary goal of constraining bar formation in galaxy 
merging, we conducted simulations with varying mass-ratios and spin 
angles $\theta_1, \theta_2$ for both disk-dominated and bulge-
dominated initial disks. The output from these simulations were 
then tested with the CNN. Since each simulation output has 200 
associated density maps (each oriented differently with respect to 
the viewing axis), we define the prevalence of bar formation as the 
fraction of these 200 images that are classified as a bar by the 
CNN. That is, we define the bar fraction as:
\begin{equation}
f_{\text{bar}} = N_{\text{bar}} / N
\label{eq:barprob}
\end{equation}
where $N_{\text{bar}}$ are the number of images detected as showing 
a bar, and $N$ is the total number of images. We also refer to this 
as the bar probability, with the terms used interchangeably 
depending on the context. We ran several disk-dominated and bulge-
dominated models, and tested the outputted data with our fully-
trained CNN to determine the prevalence of bar formation as 
functions of mass ratio $m_2$ and spin angles $\theta_1, \theta_2$. 

Our secondary goal of using a CNN to distinguish between different 
bar formation mechanisms used a new CNN with the same core 
architecture with two key differences: it was trained on a 
different set of simulated data, and utilised a different labelling 
schema (4 categories instead of 2).

\section{Results}

\begin{figure*}
\centering
\includegraphics[scale=0.24]{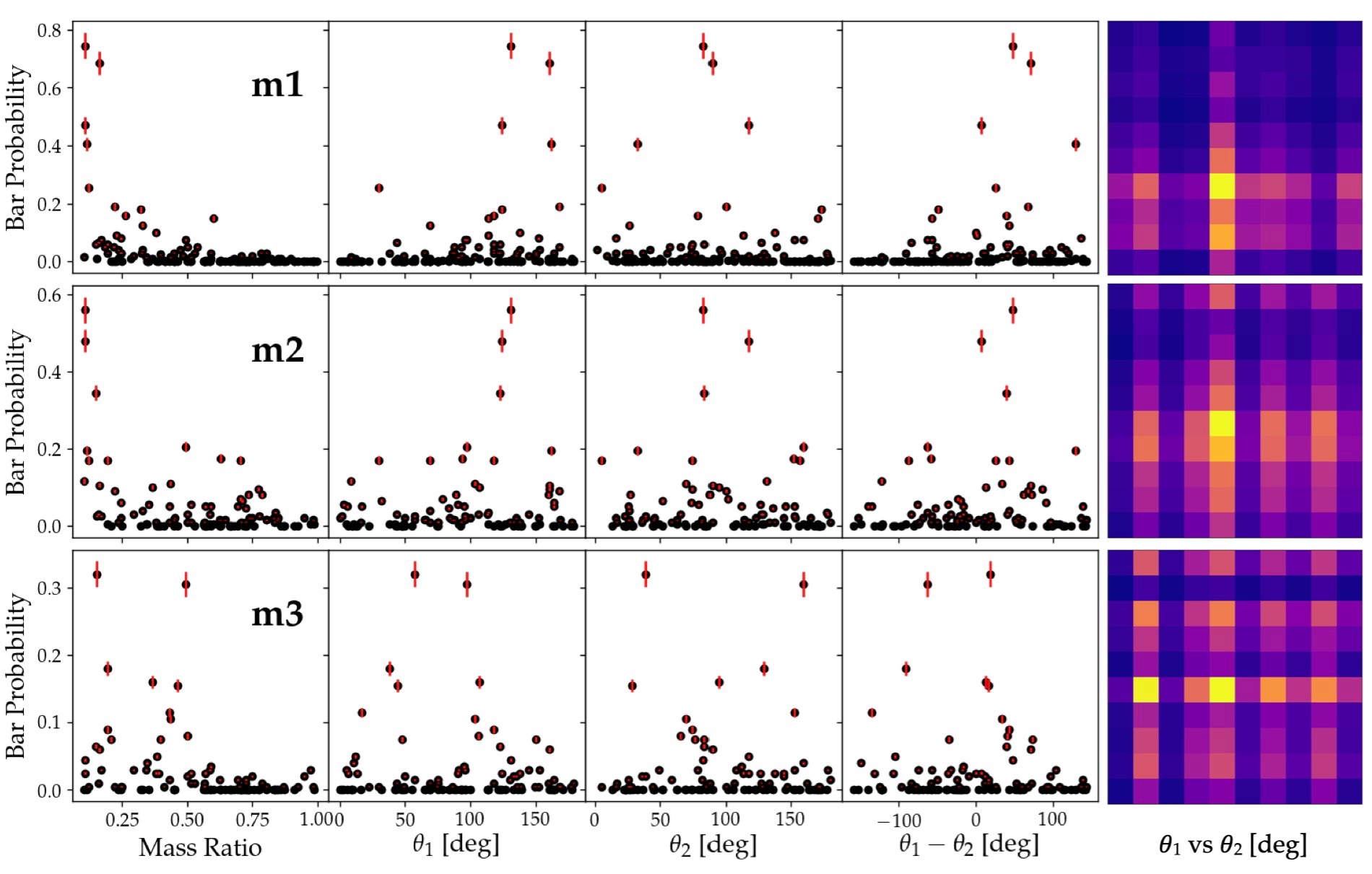} \\
\includegraphics[scale=0.24]{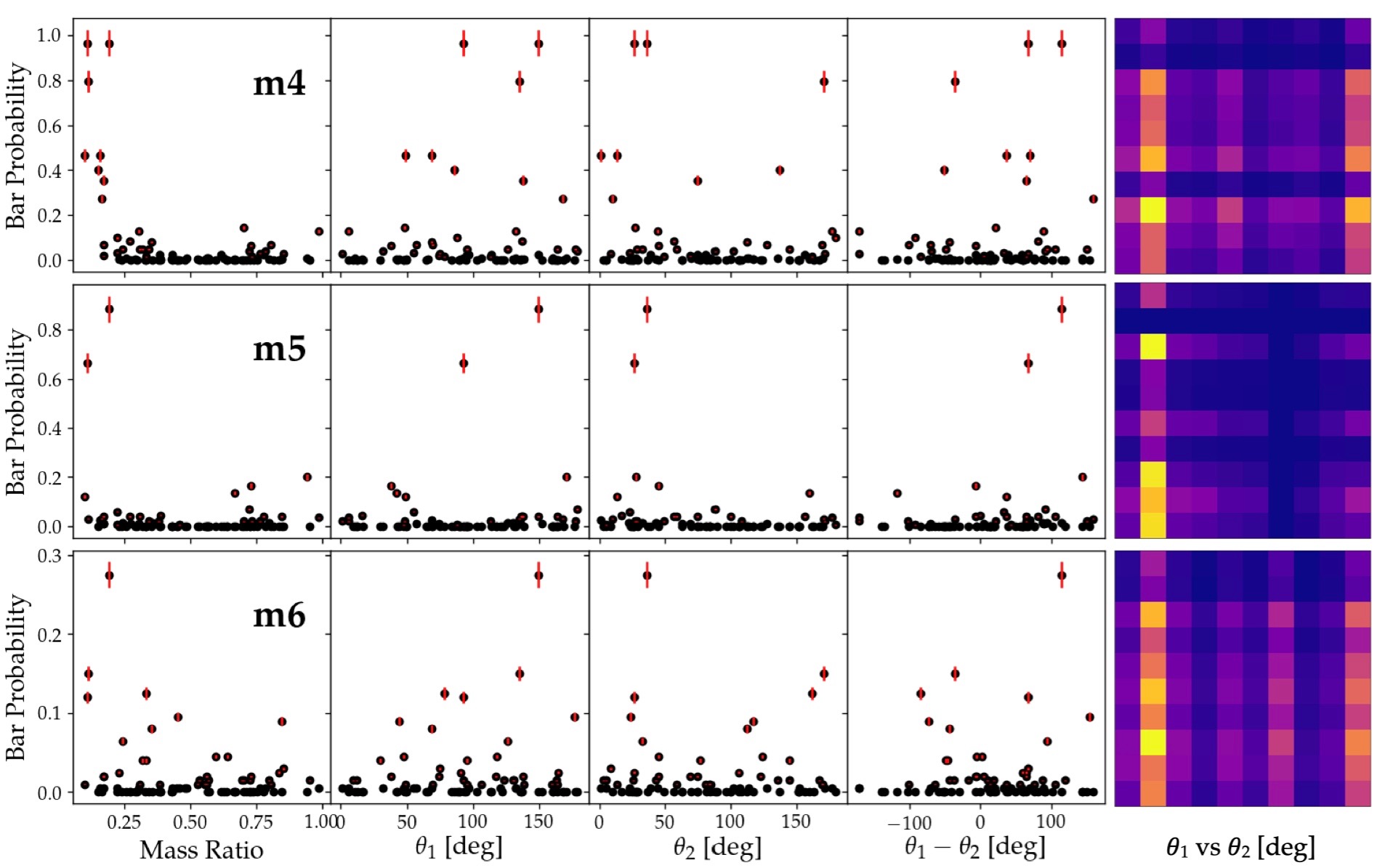}
\caption{Incidence of bar formation, expressed as the bar fraction 
$f_{\text{bar}}$ or bar probability,
as a function of mass ratio $m_2$, the difference in spin angle
$\theta_1, \theta_2$, and as a function of $\theta_1$ vs. $
\theta_2$ where $\theta_1$ is on the vertical axis.
Lighter shades in the normalised density map of $\theta_1$ vs
$\theta_2$ indicate a higher relative probability of bar 
detection.  Error bars indicate one standard error $\sigma$.}
\label{fig:m-all}
\end{figure*}

\begin{figure*}
\centering
\includegraphics[scale=0.52]{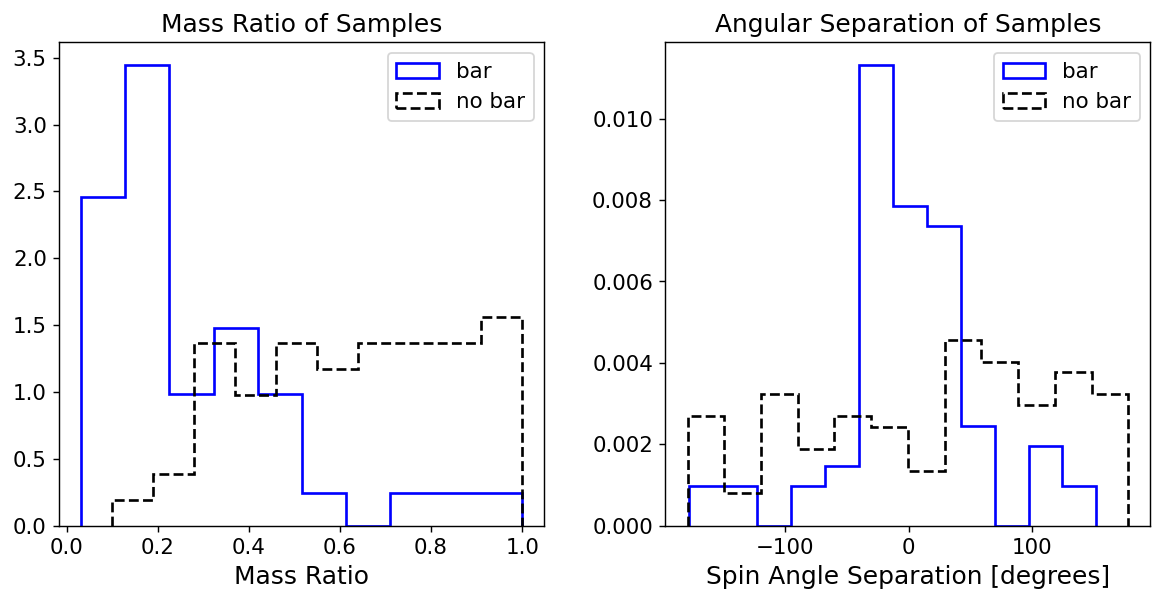}
\caption{Normalised histograms of all samples classified as either \textbf{bar} (blue) or \textbf{no-bar} (black, dotted) by our main CNN, for different mass ratios (left) and spin angle separation (right).}
\label{fig:detections}
\end{figure*}

\begin{table*}
\centering
\caption{Total bar 
fractions for the variable mass ratio $m_2 = 0.1$ to 1.0 and fixed 
mass ratios $m_2 = 0.1$ and $m_2 = 1.0$, for each of our six galaxy models.}
\label{tab:mbf}
\begin{tabular}{|c|c|c|c|c|c|c|c|}
\hline 
Model & Bar fraction & Bar fraction & Bar fraction \\ 
\ & $m_2=0.1$ to $1.0$ & $m_2=0.1$ & $m_2=1.0$ \\
\hline
m1 & 0.075 $\pm$ 0.0037 & 0.082 $\pm$ 0.0041 & 0.013 $\pm$ 0.00065\\ 
\hline 
m2 & 0.064 $\pm$ 0.0032 & 0.073 $\pm$ 0.0036 & 0.029 $\pm$ 0.0015 \\ 
\hline 
m3 & 0.048 $\pm$ 0.0024 & 0.051 $\pm$ 0.0026 & 0.03 $\pm$ 0.0015\\ 
\hline 
m4 & 0.12 $\pm$ 0.006 & 0.11 $\pm$ 0.0055 & 0.038 $\pm$ 0.0019\\ 
\hline 
m5 & 0.089 $\pm$ 0.0044 & 0.091 $\pm$ 0.0045 & 0.057 $\pm$ 0.00285 \\ 
\hline 
m6 & 0.037 $\pm$ 0.0019 & 0.037 $\pm$ 0.0019 & 0.036 $\pm$ 0.0018\\ 
\hline 
\end{tabular} 
\end{table*}

\subsection{Mass ratio}

To investigate the effects of mass ratio, we ran several hundred 
simulations for each of the six galaxy models in Table 
\ref{tab:models} for a variable mass ratio ($m_2$ between 0.1 and 
1.0) as well as for the fixed mass ratios $m_2 = 0.1$ 
(corresponding to \textit{minor merging}) and $m_2 = 1.0$ 
(corresponding to \textit{major merging}).  Each of the simulations 
were conducted with random spin angles such that $0 \leq \theta_i 
\leq 180$ degrees for $i = 1,2$.  These models are completely separate from those used to train the CNN. Instead, these models were classified by the fully-trained CNN.

Table \ref{tab:mbf} shows the detected bar fractions (i.e as 
defined in Equation \ref{eq:barprob}) for each of these three mass 
cases for our six galaxy simulation models (refer to Table 
\ref{tab:models} for their physical parameters). 
On average, the CNN detected more bars present in the 
aftermath of minor merging ($m_2$ on the order of $0.1$) as opposed 
to major merging ($m_2 \sim 1.0$).
Figure \ref{fig:m-all} gives a visualisation of the bar probability 
as functions of the parameters $m_2, \theta_1, \theta_2, \theta_1 - 
\theta_2$ and finally as a normalised probability density map 
corresponding to $\theta_1$ vs. $\theta_2$. Again we see a clear 
inverse relationship with mass ratio; lower mass ratios are more 
conducive to bar formation.

Figure \ref{fig:detections} gives a more compact visualisation of 
the overall CNN classifications, showing the normalised histograms 
of all samples classified as bar or no-bar with respect to mass 
ratio and spin angles.  As seen in Figure \ref{fig:m-all}, of those 
samples classified as bars, they generally have lower mass ratios 
and lower angular separation.  Conversely, samples classified as 
no-bar tend to have mass ratios between 0.3 and 1.0, and are 
uniform with regards to angular separation.  This latter result 
suggests that mass ratio is a more influential factor in the bar 
formation process.

It is important to stress that the 
results in Figures \ref{fig:m-all} were 
obtained by running limited sets of random parameter combinations. 
Due to time limitations, it was impossible to cover all 
combinations of $m_2, \theta_1$ and $\theta_2$, but there are 
enough data points to qualitatively infer trends and pinpoint 
certain parameter ranges that are more conducive to bar formation 
than others.

In Figure \ref{fig:m-all}, we see both reciprocal relationships 
(such as the m1 model) and a relationship with two distinct peaks 
(the m3 model). This suggests that intermediate-mass mergers may be 
just as efficient as minor mergers at producing bars for certain 
galaxy models (in the case of m3, a galaxy model with a smaller 
stellar mass).  For all models, major merging is far less conducive 
to bar formation compared to minor merging. Further simulations with a wider range of model parameters are required to definitively examine the mathematical nature of this relationship.

\subsection{Spin angles}

\begin{figure*}
\centering
\includegraphics[scale=0.22]{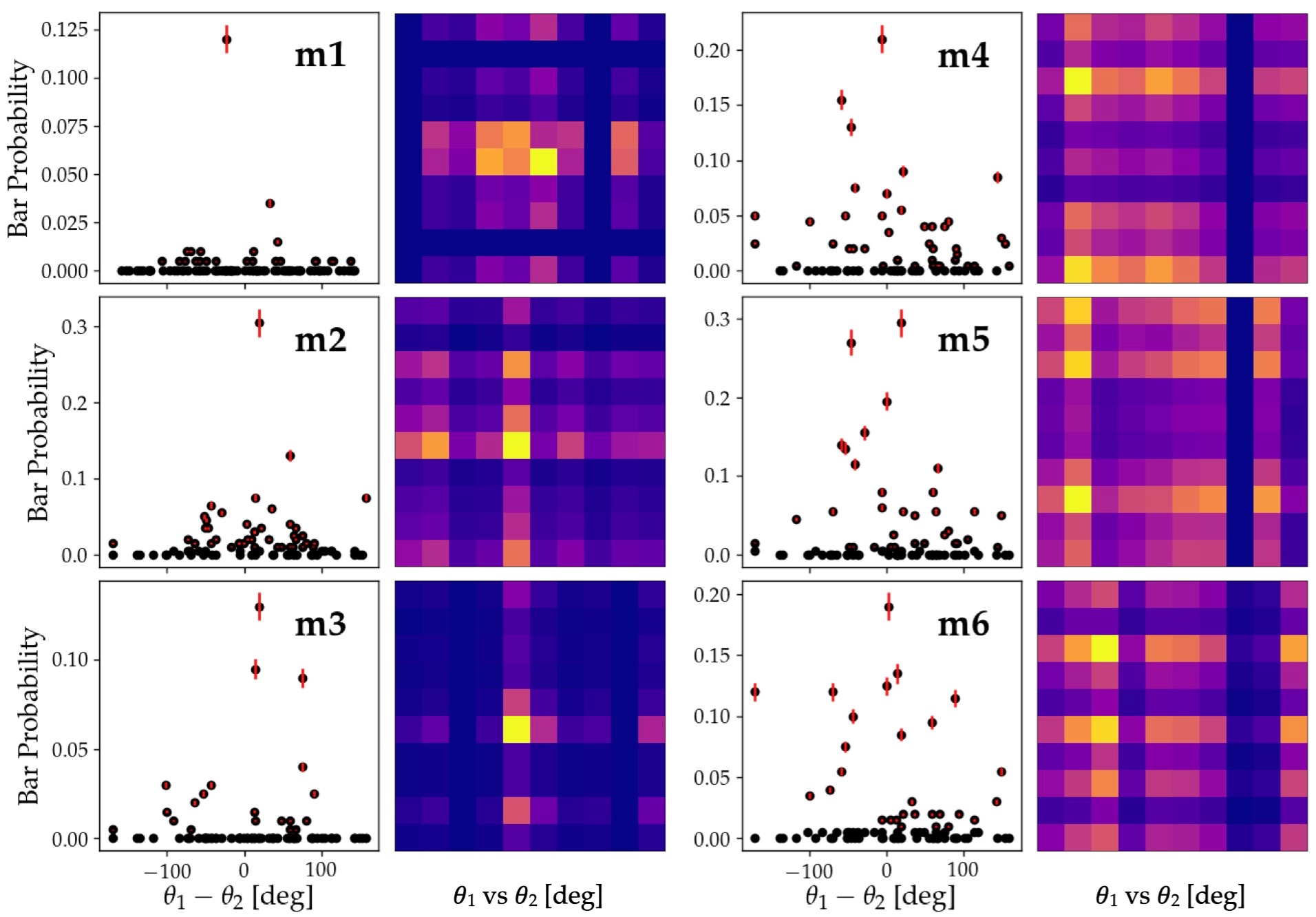}
\caption{The incidence of 
bar formation, expressed as the fraction of total images that were 
identified as bar for a given model, as a function of the 
difference in spin angles $\theta_1 - \theta_2$ and as a function 
of spin angles $\theta_1$ vs.$\theta_2$ for major mergers at a 
fixed mass ratio of 1.0.}
\label{fig:mfixed}
\end{figure*}

The m1, m2 and m3 models appear to be more strongly constrained by 
$\theta_1$, while there is a dominant $\theta_2$ constraint for the 
m4, m5 and m6 models.  Given the significant scatter in the overall 
spin angles, more simulations are needed to conclusively determine 
any specific constraints on either angle.  Since the exchange of 
angular momentum is an important factor of bar formation, it is 
better to focus on the difference in the spin angles, $\theta_1 - 
\theta_2$ (i.e how closely aligned the disks are).  We find that, for the m1, m2 and m3 models, more bars were detected for more closely aligned $\theta_1$ and $\theta_2$ (i.e $|\theta_1 - \theta_2| \approx 0$) compared to 
mergers with wildly different spin angles.  However, such a trend is not observed for the m4, m5 and m6 models in the variable-mass case.

In order to analyse these spin angle constraints independently of 
mass ratios, we ran several models for all 6 of our galaxy merger 
models with a fixed mass ratio of 1.0 (Figure \ref{fig:mfixed}), corresponding to major mergers. We previously established that 
fewer bars were detected in major merging; hence testing on major 
mergers yields a stronger constraint on the spin angles.
Again, we find that models in which $\theta_1$ and $\theta_2$ are 
more closely aligned resulted in the detection of more bars. In 
each case, the parameter with the highest fraction of bar detection 
is located very close to $\theta_1 - \theta_2 = 0$. This is also true for the m4, m5 and m6 models. Importantly, 
these results were consistent across our various models despite 
their differences in bulge-to-disk ratios and dark matter contents.

\section{Discussion}

To verify these constraints, we first identified several key 
cases that involved contrasting two simulation models. The most 
important of these test cases is to compare a major merger with a 
minor merger. We did this for the small-bulge m1 disk model (Figure 
\ref{fig:maj-min}) as well the large-bulge m4 model with a bulge-
to-disk ratio of 0.5 (Figure \ref{fig:bulge-maj-min}).  We decided 
to use the m1 model instead of the bulge-less m6 model, and the m4 
model instead of the 1:1 bulge-to-disk m5 model, in order to keep 
in line with more realistic MW-type galaxies.

Since our models in Figures \ref{fig:m-all} and \ref{fig:mfixed} 
found that mergers with low angular separation were more conducive 
to bar formation, we also analysed a major merger for equal spin 
angles $\theta_1 = \theta_2 = 0$.  The latter of these 
cases is highly unrealistic given how statistically unlikely it is for two 
prograde galaxies to merge with the same orientation.  We also 
compared a simulation of the m4 model in the isolated and minor 
merger case.

\subsection{The bar formation process}

\begin{figure*}
\centering
\includegraphics[scale=0.16]{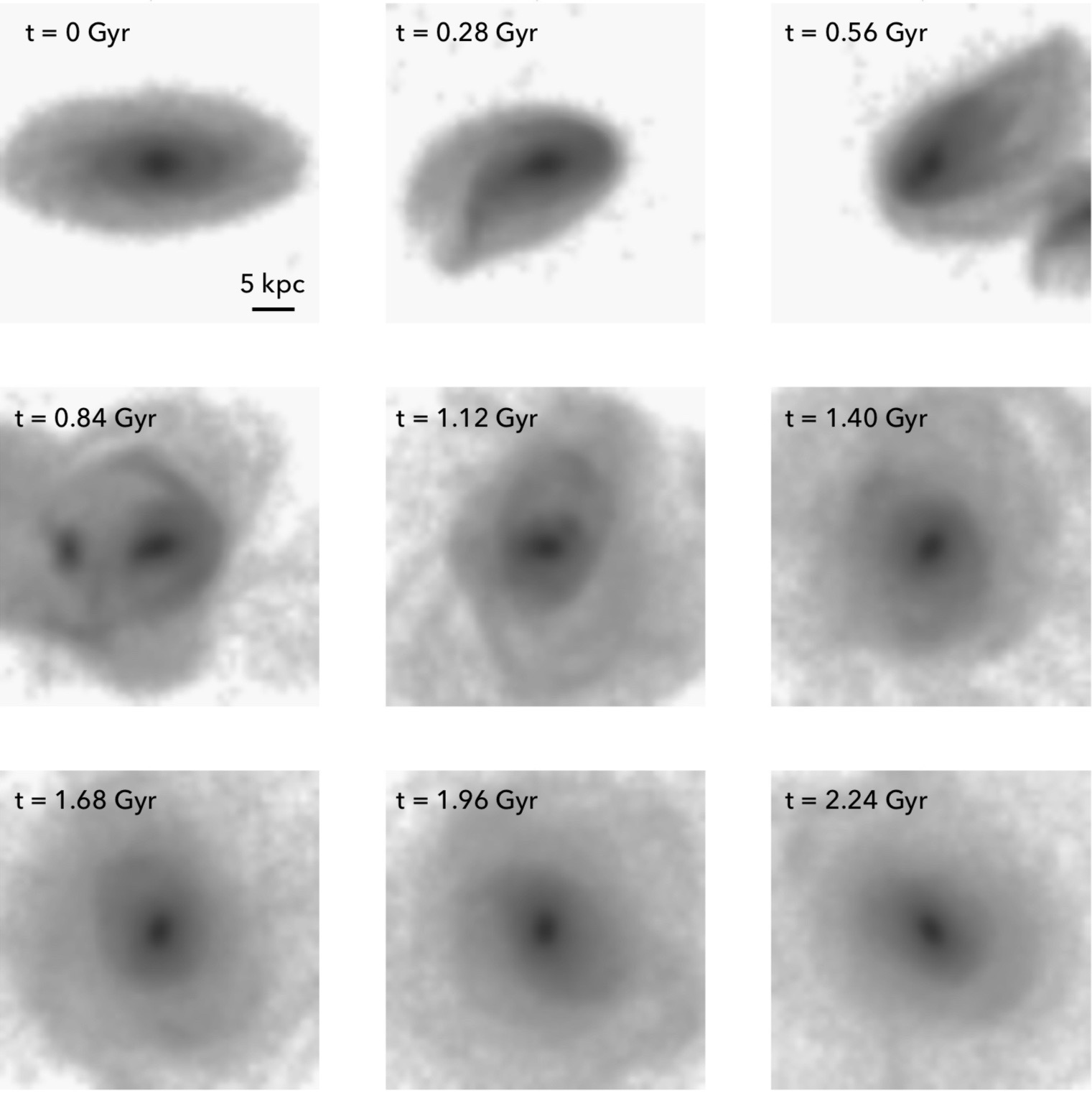}\medskip

\noindent\hrulefill\bigskip

\includegraphics[scale=0.16]{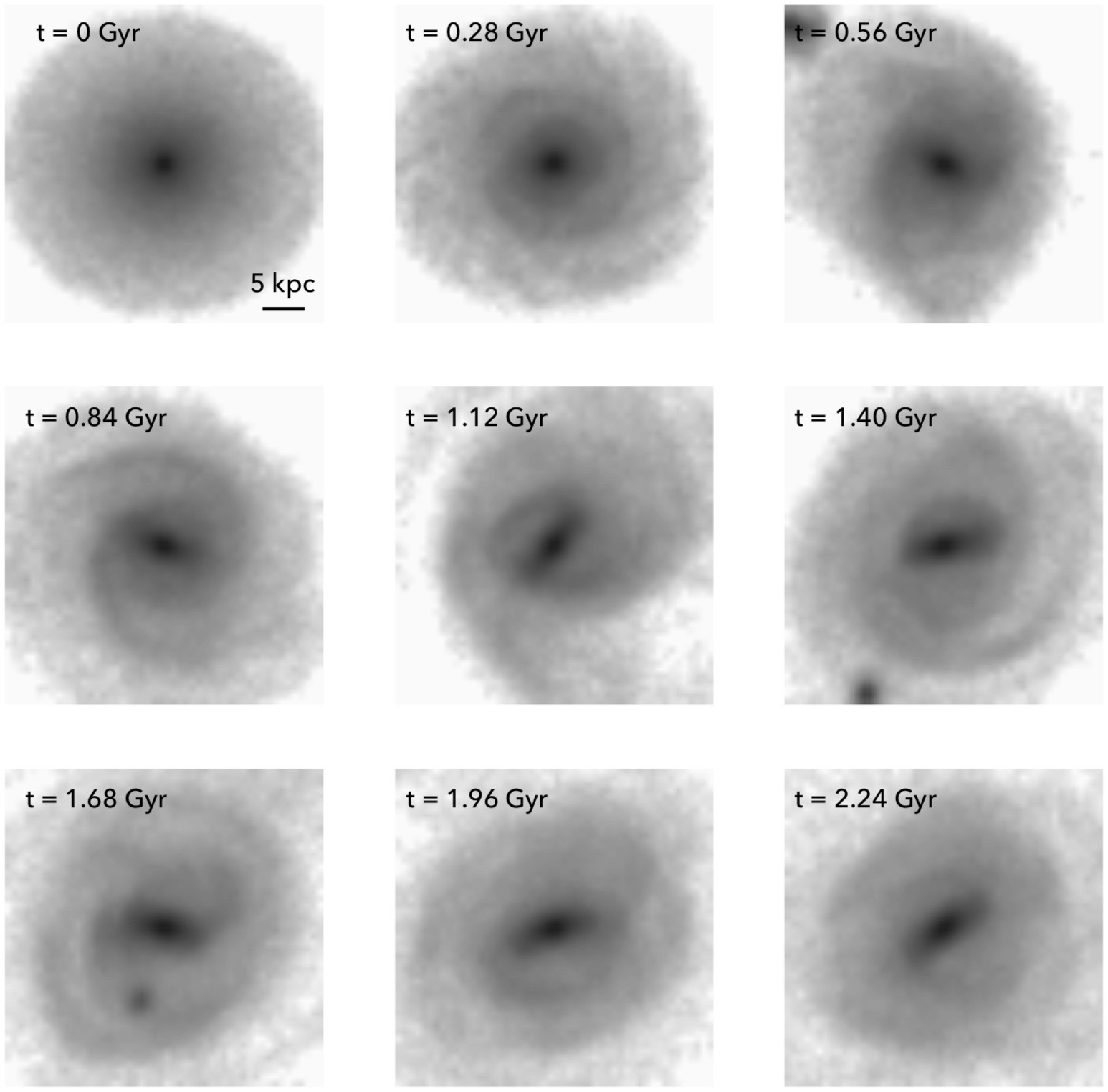}
\caption{Time evolution of the 2D normalised mass density 
distribution ($x$-$y$ plane projection) for a major merger with 
mass ratio $m_2 = 1.0$ (top) and minor merger with mass ratio
$m_2 = 0.1$ (bottom) for the small-bulge m1 model. Each frame 
measures 35 kpc by 35 kpc (these dimensions are used throughout) 
and each timestep is 240 Myr.}
\label{fig:maj-min}
\end{figure*}

It is clear from Figure \ref{fig:maj-min} that a stellar bar is 
readily formed in the minor merger case, while no bar is produced 
in the major merger.  Importantly, the major and minor merger cases 
are distinguished by the timescale of the merger.  In the major 
merger case, the galaxies merge destructively between 840 Myr and 
1.12 Gyr, while the minor merger completes between 1.68 Gyr and 
1.96 Gyr (some 800 Myr later).  Our simulation results in Figure 
\ref{fig:m-all} have shown that minor merging is more conducive to 
bar formation.  This is important considering the well-studied 
effects of minor merging in the overall evolution of galaxies 
\citep{reichard-2008}.  Simulations have also shown the vast 
majority of all mergers in the Universe to be minor mergers 
\citep{lotz-2010}.  It is important, however, to keep in mind the 
low overall bar detections in Table \ref{tab:mbf}; bar formation in 
galaxy merging is rare.

In most cases where a bar is detected in the aftermath of the 
merger event, the bar is first formed due to 
tidal interaction with the approaching galaxy, after which it then 
survives the merger process. In the case of major merging, the 
merger is destructive. Conversely, minor merging is constructive.  
This is sensible given the difference in gravitational effects. 
Smaller companions tend to slowly spiral into their parent (helping 
to induce a tidal bar), and by the time the merger is complete they 
are usually sufficiently stripped of mass so as to not warp or 
disturb the parent disk.  Equal mass mergers are more destructive 
due to the quick head-on collision (such as what we see in Figure 
\ref{fig:maj-min}) that severely warps the stellar disk.

\begin{figure*}
\centering
\includegraphics[scale=0.16]{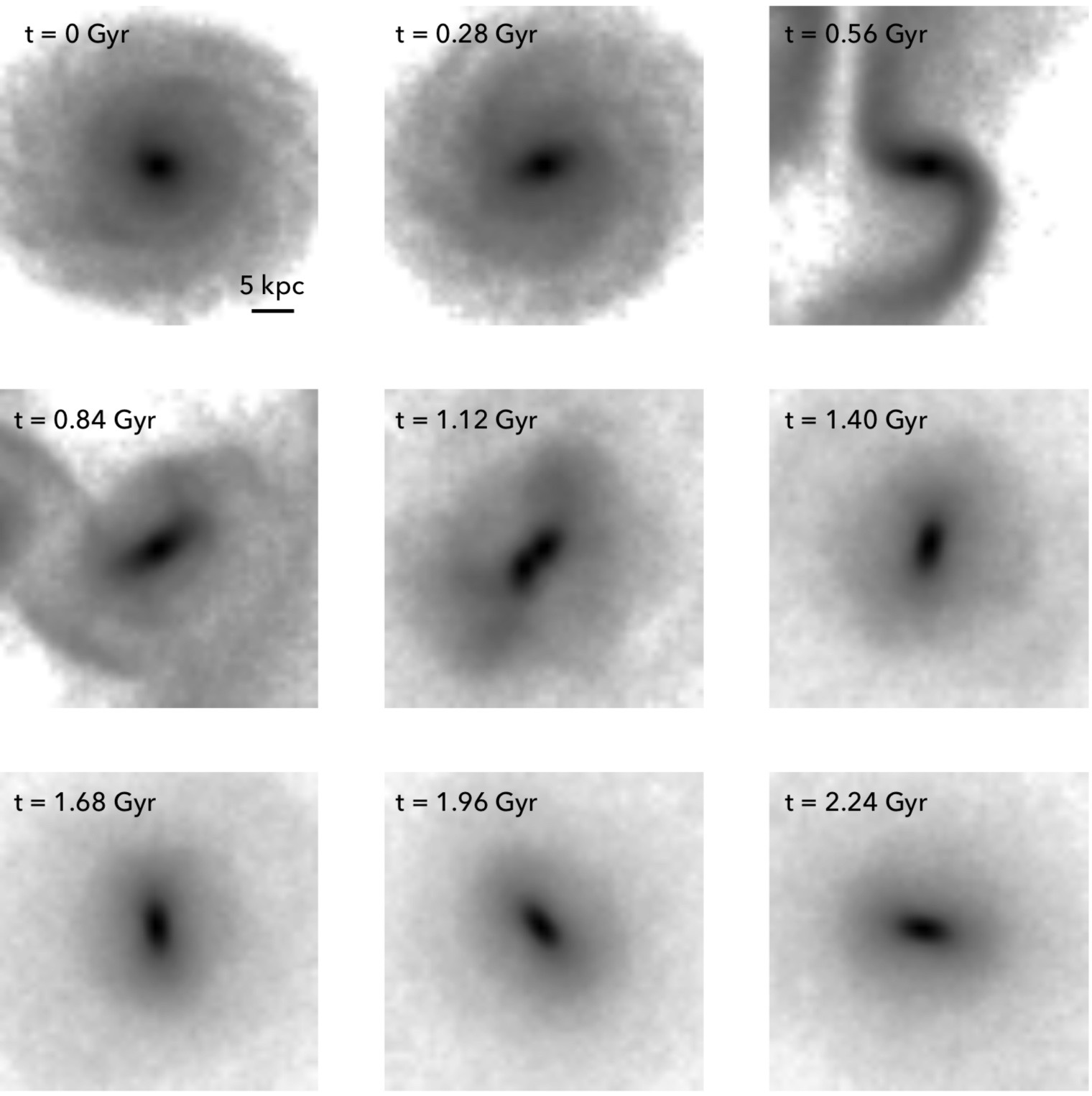}
\caption{Time evolution of the 2D normalised mass density 
distributions for our simulations of a major merger for $\theta_1, 
\theta_2 = 0$, demonstrating the formation of a bar in a major 
merger with $\theta_1 = \theta_2 = 0$ degrees.  Frame dimensions 
are as in Figure \ref{fig:maj-min}}
\label{fig:majmer-eqth}
\end{figure*}

It is this latter notion of whether a bar survives or is destroyed 
that is they key motivation behind finding constraints for our 
simulation parameters. These constraints will help guide further 
investigation into the exact mechanisms with which bars can survive 
a galaxy merger unscathed.  It is important to note that there are 
two mechanisms at play here; the initial pre-merger tidal 
interactions, and the actual merger itself.  Previous studies, e.g 
\citep{peirani-2009, di-matteo-2010}, have shown that bars can form 
due to the tidal interactions before a merger, however they did not 
consider what happens to the bar during the merger itself. To 
better illustrate this dual-nature, Figure \ref{fig:majmer-eqth} 
shows a major merger with $\theta_1 = \theta_2 = 0$ degrees.

First and foremost, Figure \ref{fig:majmer-eqth} shows
bar formation in the major merger case. 
So although the results of Figure \ref{fig:m-all} show 
that minor merging is much more promising, we have demonstrated 
that bar formation is still possible for major mergers.
The spin angles (i.e the orientation of the galaxies) are also key 
to whether or not a bar is produced.  In Figure
\ref{fig:majmer-eqth}, we see that a bar is induced at around 0.56 
Gyr. The actual merger takes place between 0.84 and 1.12 Gyr.
However, the snapshot at 1.12 Gyr shows the bar 
has buckled into two loosely connected bulges (likely due to the 
force of the merger), but by 1.40 Gyr the stellar bar has reformed 
and continues to exist for the rest of the simulation timescale.  
This is strong evidence for the role of spin angles in the 
formation and survival of stellar bars in merging.  So while it is 
possible for bars to be induced prior to the merger due to tidal 
interactions, regardless of orientation (as tidal interactions are 
gravitational in nature), orientation is key to whether a bar can 
be reformed in the aftermath of a galaxy merger.  It is this latter 
mechanism that we wish to highlight in this current study. 

Angular momentum is a key factor in galaxy merging
\citep{athana-2005, pedrosa-2015}.
The results of Figures \ref{fig:mfixed} and Figure 
\ref{fig:majmer-eqth} show that major mergers with more closely 
aligned orientations (i.e $\theta_1-\theta_2| = 0$) have higher bar 
detection rates.  This is likely due to a more favourable transfer 
of angular momentum that, as Figure \ref{fig:majmer-eqth} shows, is 
sufficient to regenerate the stellar bar.

Thus we have shown that there are two distinct phases that govern 
the overall bar formation process in galaxy merging; (1) pre-merger 
tidal interactions, and (2) reformation of the bar during and/or 
after the merger.  Our results have shown that the minor merging of 
two galaxies with near-identical spin angles is most conducive to 
bar formation in the aftermath of a merger.  This highlights the 
importance of angular momentum transfers, and also the role of 
gravitational forces given the sharp decline in bar 
detections as the mass ratio increases.  We expect that the second phase is 
more dominant in major merging due to the destructive nature of the 
merger whereby strong gravitational forces greatly perturb the 
stellar disk.

\subsection{Effect of stellar bulges}

\begin{figure*}
\centering
\includegraphics[scale=0.16]{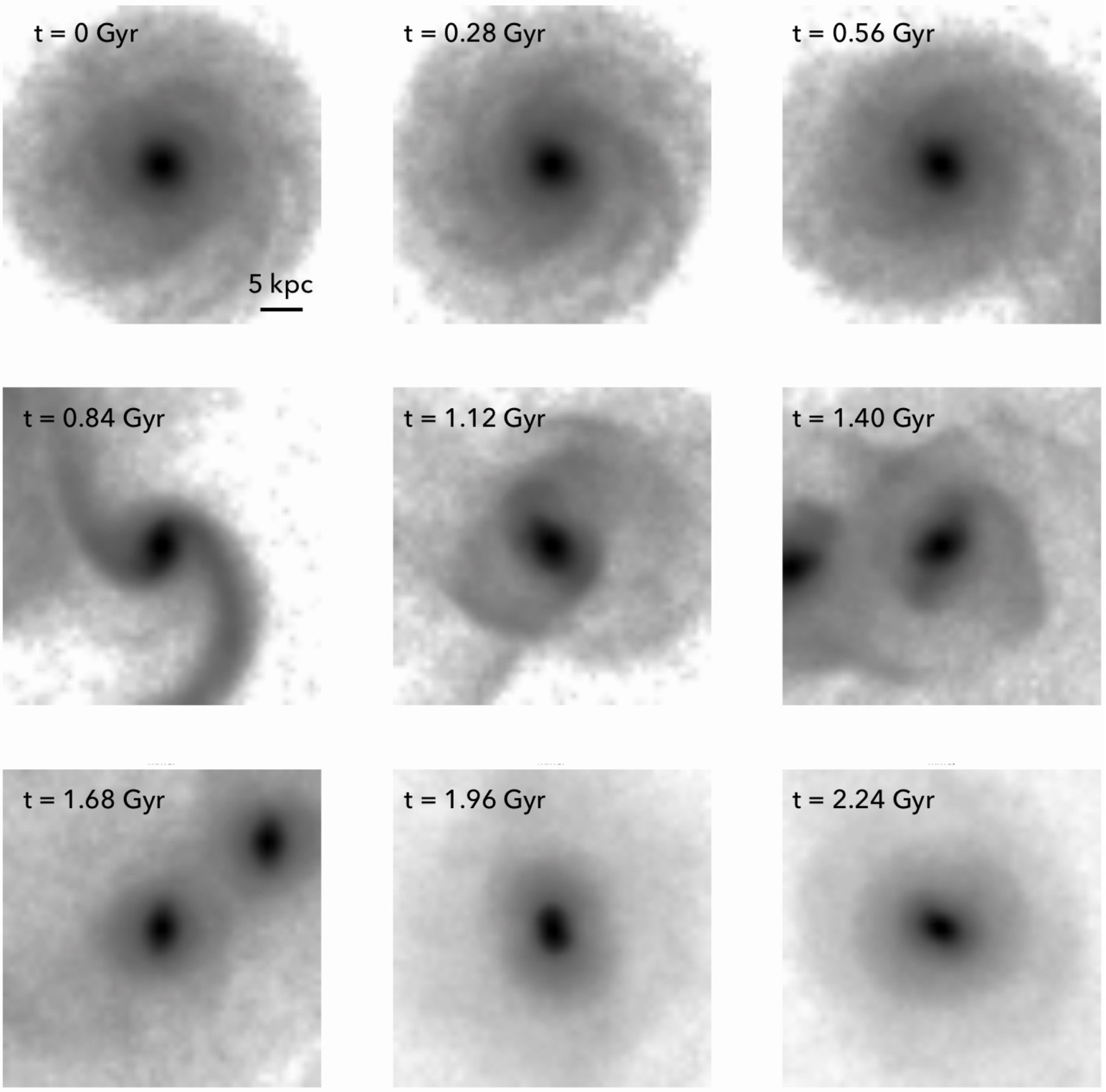}\medskip

\noindent\hrulefill\medskip

\includegraphics[scale=0.16]{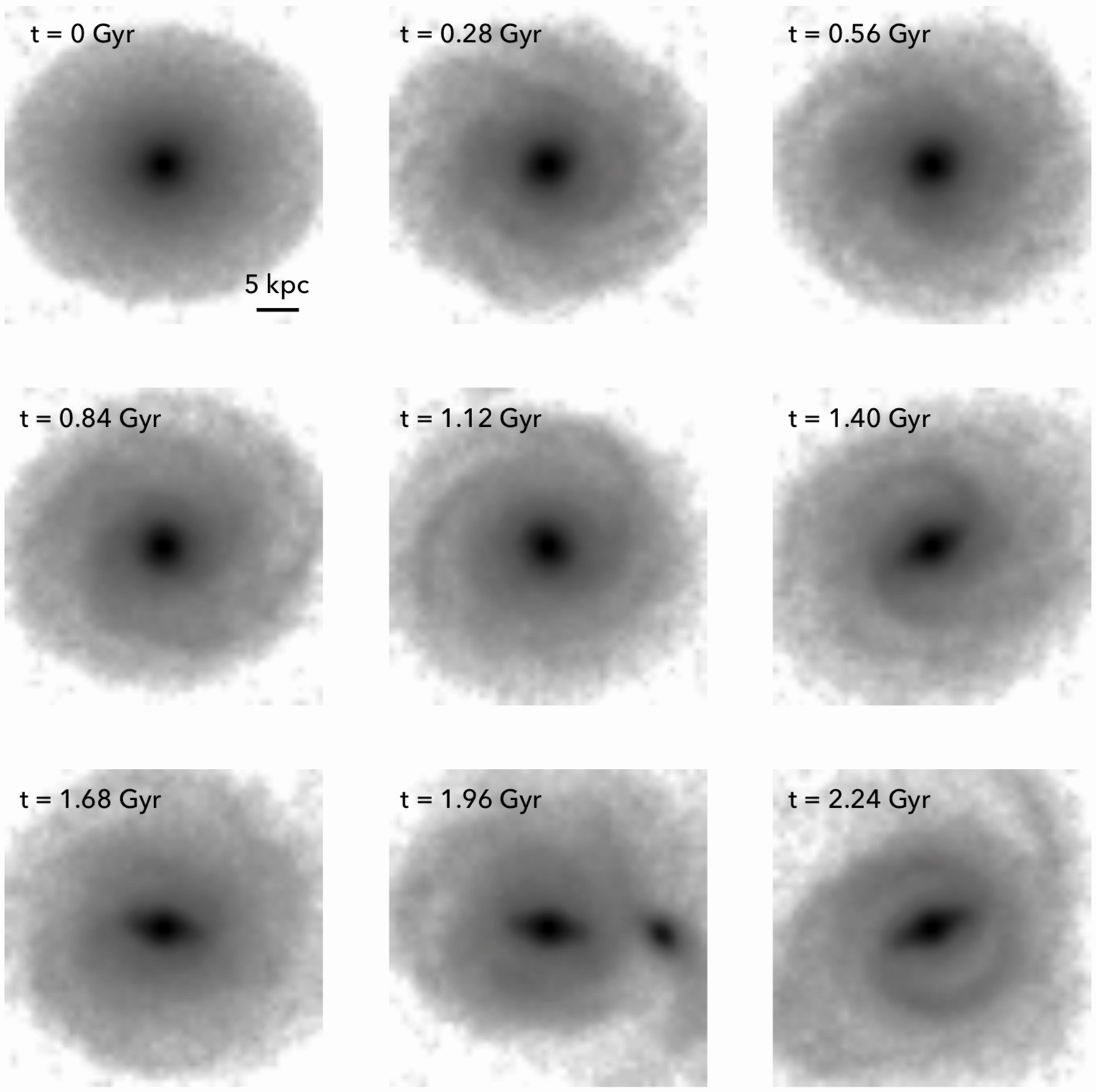}
\caption{Time evolution of our simulations of 
a major merger with mass ratio $m_2 = 1.0$ (top) and a minor 
merger with mass ratio $m_1 = 0.1$ for a model with a bulge-to-disk 
ratio of 0.5 (bottom).}
\label{fig:bulge-maj-min}
\end{figure*}

The bar fractions in Table \ref{tab:mbf} are lower for models with 
smaller bulges. This is somewhat counter-intuitive given that 
bulges tend to suppress bar formation in isolation. This suggests 
that a bulges may play a more important role in galaxy merging.  To 
visualise whether stellar bulges have a meaningful impact on bar 
formation, we ran a test simulation with the m4 model (bulge-to-
disk ratio of 0.5) for both a major and minor merger (see Figure 
\ref{fig:bulge-maj-min}).  As anticipated (based on the results of 
Figure \ref{fig:m-all} and the bar fractions in Table 
\ref{tab:mbf}), major merging is destructive while minor merging is 
constructive. A key physical observation is that the addition of a 
sizeable stellar bulge has increased the time taken for a bar to be 
induced.  In the minor merging case, a bar is only induced after 
around 1.68 to 1.96 Gyr (considerably more time than in Figure 
\ref{fig:maj-min}).  This supports the well-known dynamical role 
that stellar bulges play in stabilising disk instabilities 
\citep{kataria-2017} and hence inhibiting bar growth 
\citep{sellwood-1993}.
 
\subsection{Comparison to the isolated case}

\begin{figure*}
\centering
\includegraphics[scale=0.16]{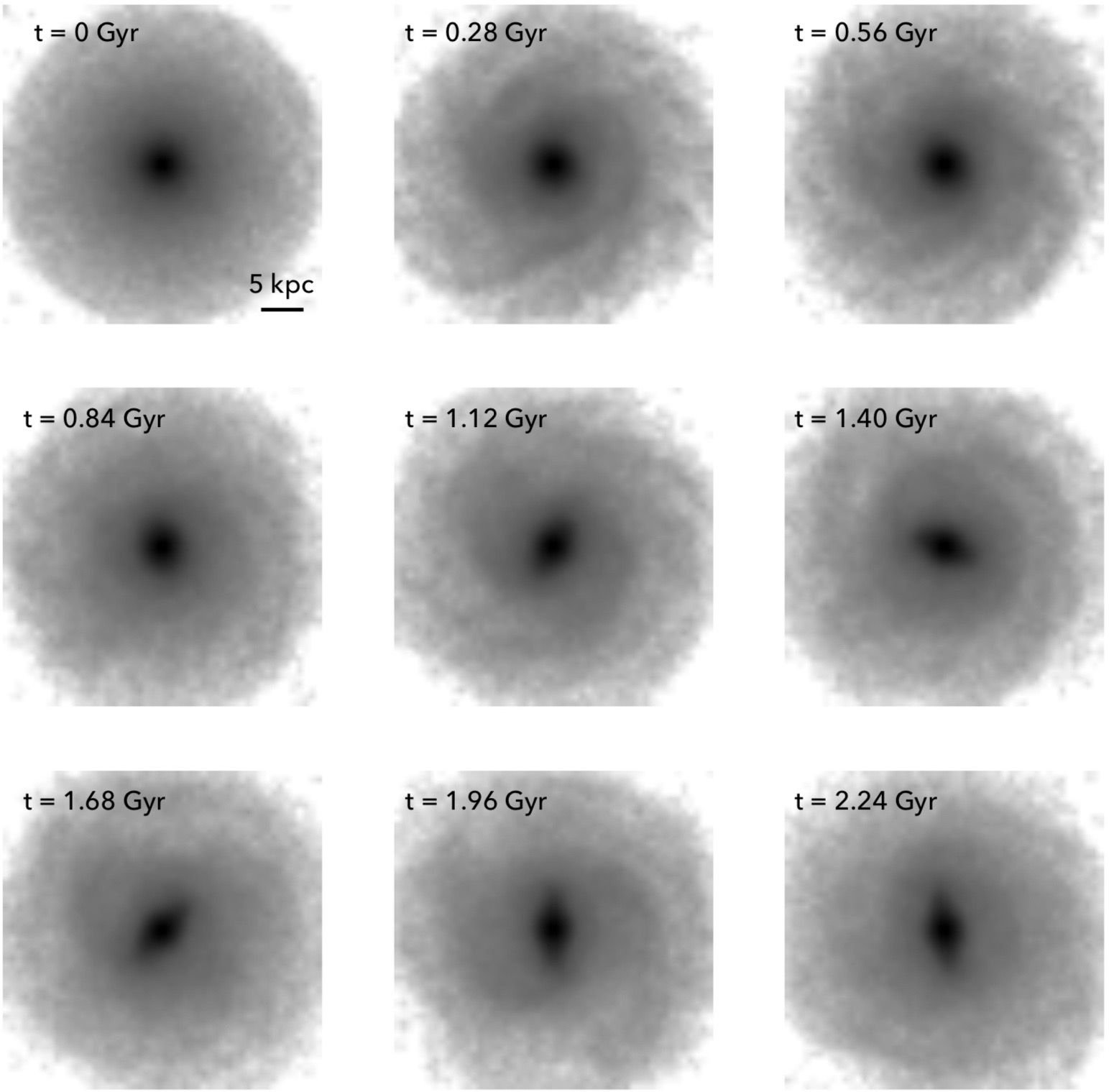}\medskip

\noindent\hrulefill\medskip

\includegraphics[scale=0.16]{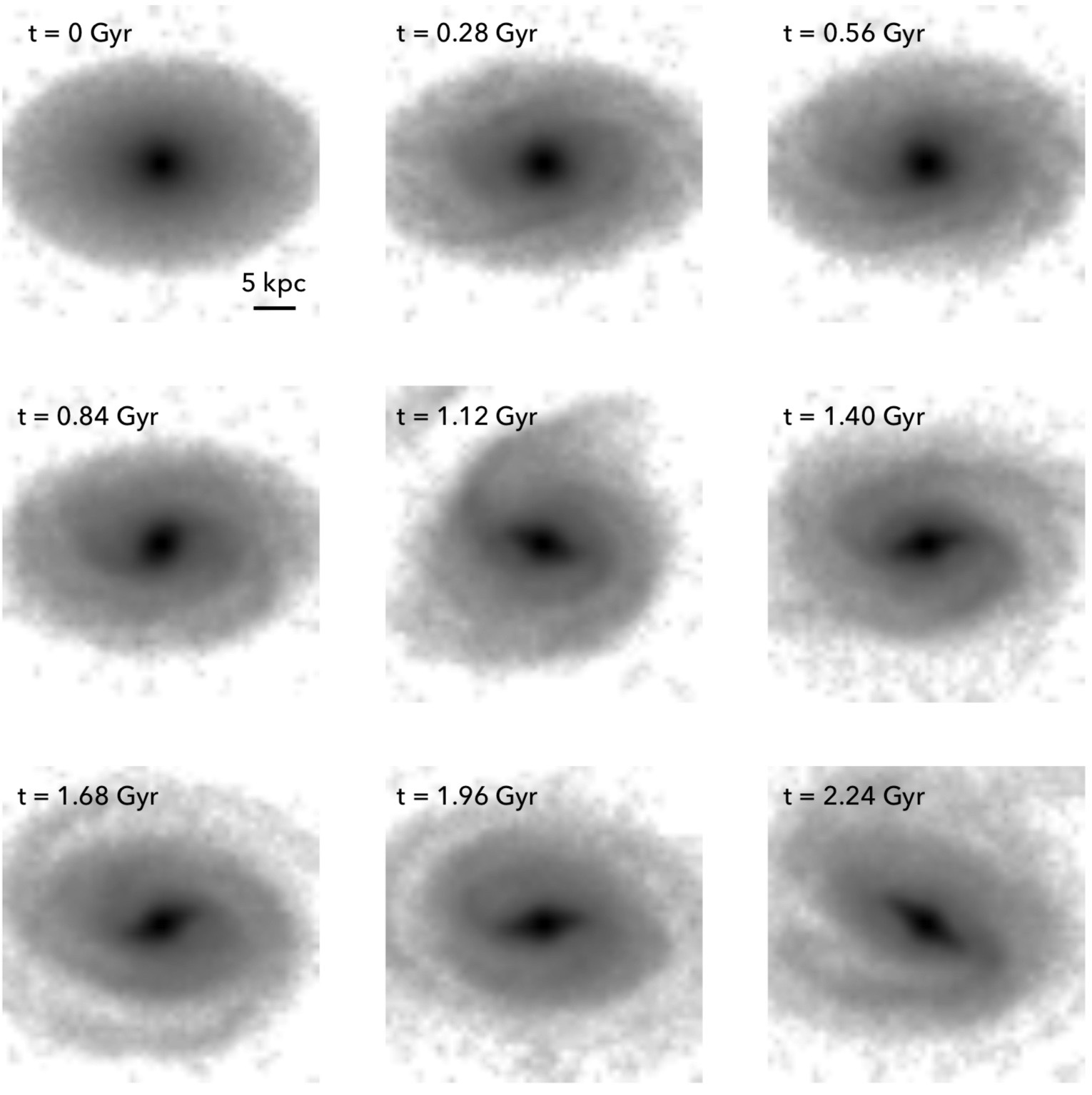}
\caption{Time evolution of our simulations of 
a galaxy with a bulge-to-disk ratio of 0.5 in the isolated case 
(left) and merger case (right).}
\label{fig:bulge-iso-mer}
\end{figure*}

To better understand the role of minor merging in bar formation, we 
ran a simulation of the m4 galaxy model in both the isolated case 
and for a minor merger with mass ratio $m_2 = 0.1$ (see Figure 
\ref{fig:bulge-iso-mer}).  The idea is that the bulge should 
suppress bar formation, yet in the minor merging case of Figure 
\ref{fig:bulge-iso-mer}, we see that bar is induced as early as 
1.12 Gyr, while there is only evidence of a weak bar in the 
isolated case from roughly 1.96 Gyr onwards.  The bar formed in the 
minor merger is more well defined and is also accompanied by spiral 
arms.

That minor merging can induce 
bars in bulge-dominant disks is 
particularly important since it thought that bulges can stabilise 
disk instabilities \citep{binney-2008}, particularly in isolation. 
\citet{kataria-2017} investigated upper limits on bulge-to-disk 
ratios for bar formation 
in isolation, finding that massive bulges stabilise the disk by 
cutting off angular momentum exchange between the disk and halo.

\citet{lotz-2010} investigated the role of mass ratios in galaxy 
morphology with cosmological simulation, and notes that the vast 
majority of galaxy mergers will 
be minor mergers rather than equal-mass major merging. In terms of 
morphological evolution, minor merging can lead to morphological 
distortion \citep{reichard-2008} that ultimately affects galaxy 
properties \citep{darg-2010}. Minor merging is a crucial process 
in the formation and evolution of galaxies. We have found that 
mass ratio indeed plays a key role in bar formation, confirming the 
importance of mass ratio when studying the processes that govern 
galaxy merger.

\subsection{Pattern speeds}

One important factor to ensure the accuracy of the classifications  
is to determine whether the bars 
as seen in Figures \ref{fig:maj-min}, \ref{fig:bulge-maj-min} and 
\ref{fig:bulge-iso-mer} are actually real bars, or instead just 
elongated bulges. 
One such method is to analyse the pattern speed $\Omega_b$,
i.e the rate (or frequency) with which the bar 
rotates. The pattern speed is a very important parameter when it 
comes to the dynamics of barred galaxies
\citep{sellwood-1993, sellwood-1998, athana-2005}. This is since 
rotation is a key physical feature of bars \citep{sellwood-1993} 
that distinguishes them from elongated bulges or other irregular 
morphological features.
There are many well established methods to analyse pattern speeds 
\citep{gerssen-2003, aguerri-2015, wuyuting-2018} including the 
classic kinematic-based approach of \citet{tremaine-1984} (also 
known as the TW method).  The TW method is heavily reliant on 
additional data such as line-of-sight velocity maps and is more 
suited to observational data (see \citep{aguerri-2015}).

\section{Classifying bar formation mechanisms}

\begin{figure*}
\centering
\includegraphics[scale=0.25]{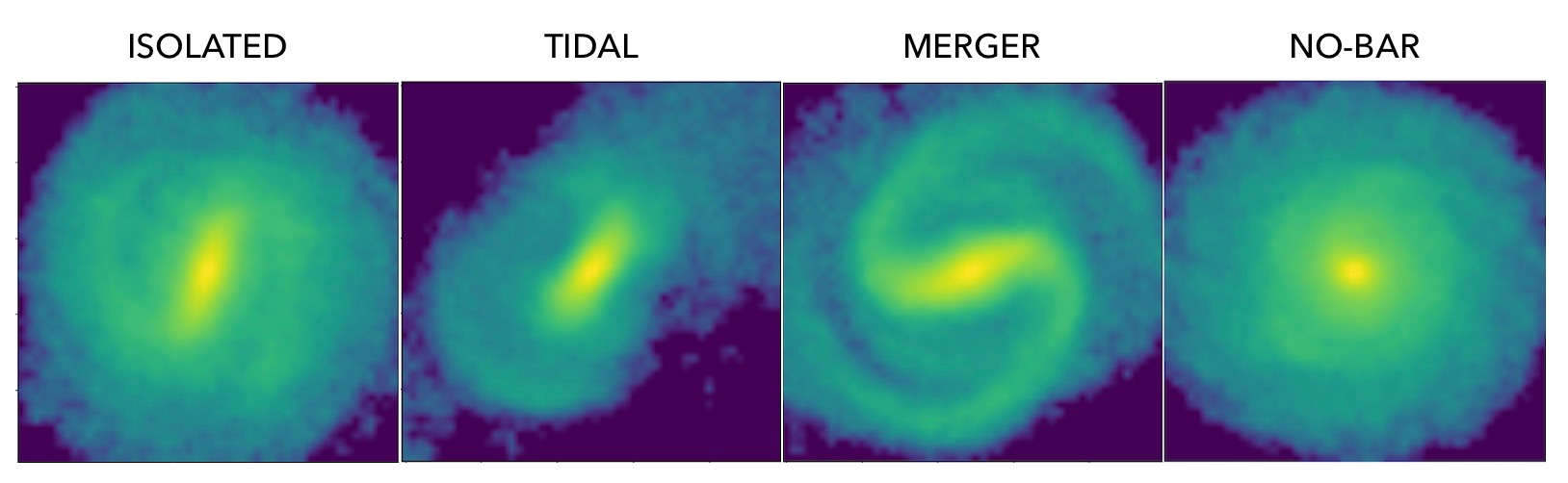}
\caption{Images in each of the four different morphological 
categories considered for our secondary, bar-type CNN}
\label{fig:4bar}
\end{figure*}

\begin{figure}[h!]
\centering
\includegraphics[scale=0.5]{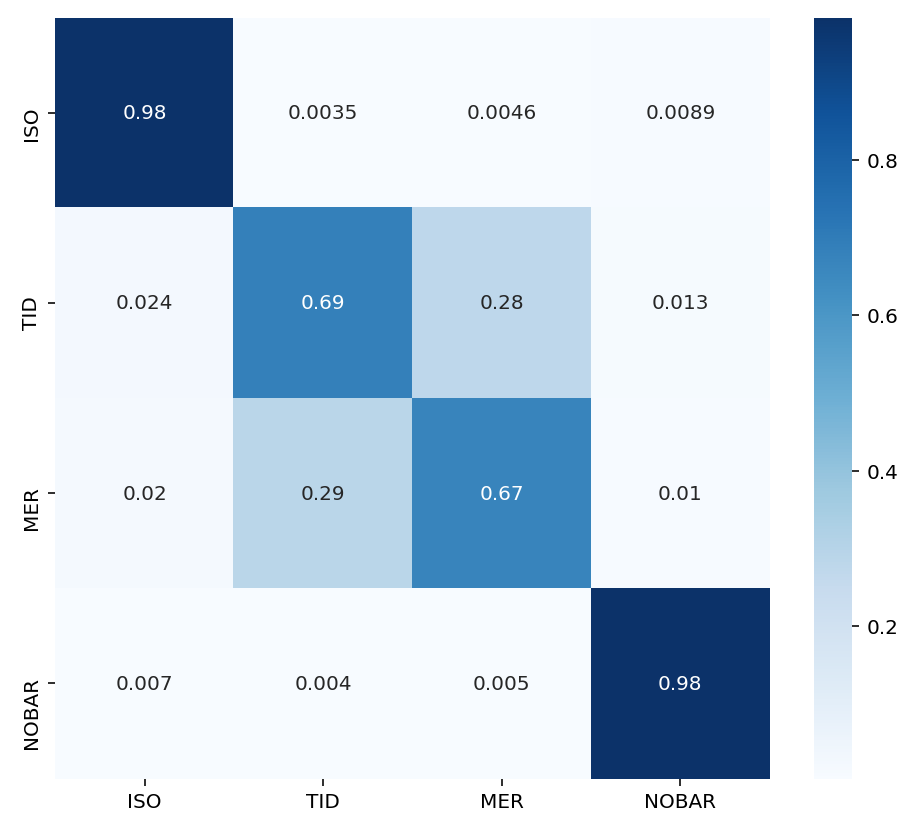}
\label{fig:confusion}
\caption{The confusion matrix of classification accuracies for our CNN. Here the columns represent the actual category and the rows represent the predicted category.}
\end{figure}

In this section, we discuss our secondary aim of classifying an 
image of a barred galaxy according to the mechanism with which it 
was formed. As this is a separate CNN from our main CNN, we will refer to it as the ``bar-type'' CNN.
Figure \ref{fig:4bar} illustrates the rationale for 
this aim.  It shows images of galaxies from our simulations in the 
isolated, tidal and minor merger cases, as well as an image with no 
bar.  Just as the human eye can discern between these four cases, 
we wish to determine whether a CNN can do the same.  For this step, 
we use the same CNN architecture used for our main analysis, albeit 
instead of 2 nodes in the output layer there are now 4 
(corresponding to the 4 categories). This new network was trained 
with a much smaller training set of 4800 images obtained from 
running the m1, m2 and m3 models for the isolated, tidal and merger 
case (with images that showed no bar formation added to the ``no-
bar'' category). This smaller set was visually classified up to an 
accuracy of 100\% with both observers agreeing on all 
classifications, however it is likely a larger set would be subject 
to more uncertainty. We randomly partitioned this training set such 
that 80\% of the samples were used the train the network, with the 
remaining 20\% used for validation purposes.  The network was 
trained up to maximum total accuracy over 300 epochs with the same 
internal architecture as the main CNN albeit it with four output 
nodes instead of two.

One way to determine how accurate the CNN is at distinguishing 
between multiple categories is by looking at the confusion matrix 
(a.k.a error matrix).  The confusion matrix lists the actual 
categories and predicted categories of a set of samples.  Assuming 
an 100\% accurate network, the confusion matrix should be a 
diagonal matrix.

As can be seen from the confusion matrix of our bar-type
CNN in Figure 
\ref{fig:confusion}, there were mixed results. While testing 
accuracies for the isolated bar and no-bar case were consistently 
above 98\%, the accuracy dropped off to 68\% and 67\% for the 
tidal bars and merger bars respectively (Figure 
\ref{fig:confusion}). This is likely due to their morphological 
similarities (see Figure \ref{fig:4bar}).
Since isolated bar formation is generally not 
accompanied by morphological features such as tidal tails or 
otherwise distorted disks, it is much easier to distinguish between 
both isolated bars and tidal bars, and isolated bars and merger 
bars. Comparing tidal bars and merger bars is not as 
straightforward since both share significant morphological 
distortion. This can be seen in the values in the confusion matrix 
(Figure \ref{fig:confusion}) where the rates of 
misclassification are well below 1\% for the other 3 cases, and 
also why tidal bars were classified as merger (and vice versa) with 
accuracies of 29\% (and 28\%). 

An alternative method to discriminate between tidal and merger bars 
may be to compare the kinematics as given by 2D maps of the line-
of-sight velocity dispersion. Neural networks trained 
on kinematic data may more accurately distinguish between tidal and 
merger bars. This is since the velocity dispersion is much more 
uniform in tidally barred galaxies, since the structure of the 
galaxy remains for the most part intact, while for merger galaxies 
there will be significant distortion and irregularities in the 
velocity dispersion due to the impact of the collision. So 
although the normalised mass distribution may show similar 
morphological features in tidal and merger bars, the differences 
may only become apparent when considering the kinematic data. This 
is a promising consideration for future studies.

\subsection{Use of neural networks}

As our analysis is heavily reliant on the use of neural networks, 
it was important to ensure that the networks were accurate. In 
general, the performance and accuracy of neural networks is
largely determined by both the quantity and
quality of the training set. Memory limitations on the Pleiades 
cluster necessarily impose an upper limit on the amount of data 
that we can train with, while limits on the visual classification process introduce an inherent uncertainty in the training set's accuracy.

We found evidence of misclassification occurring with some major mergers where, although the final image was classified as a bar, in 
reality the image was of a non-rotating, prolate bulge (an example of this situation can be seen in the major merger result of Figure 
\ref{fig:maj-min}). One of the key limitations of this deep learning approach is that the classification of galaxies is based 
on a single, static image; hence ignoring the need for bars to 
rotate.

An important issue arose with images showing near edge-on galaxies. 
Here the corresponding probability approached 
50-50 bar and no-bar, i.e a random classification. 
A hasty solution to this may well be to restrict the outputs to 
solely face-on orientations with rotations only, however to do so 
would severely limit the usefulness of the CNN and its ability to 
mirror real-world imagery. Instead, a combination of a higher 
resolution for the input image, coupled with a much larger training 
set, will likely result in improved performance at high 
inclinations. The case of perfectly edge-on galaxies will 
likely still remain close to random as, especially for the case of 
isolated bars where there is little morphological distortion, there 
is simply not enough information in the image for the various 
convolution layers to work with. Despite this, it may be feasible 
to classify bars according to their peanut or X-shape 
\citep{raha-1991, perez-2017} when viewed edge-on.
We also encountered a case of incorrect classification in the case 
of an image of a spiral arm with no stellar bar, Although this 
was only observed in one case out of the hundreds of visually 
verified cases, it is still an issue worth noting. It may be 
possible that the CNN confused the presence of spiral arms with 
that of a bar. The solution to this would be to explicitly include 
spiral-only samples labelled as ``no-bar'' in the training set.

It is also important to consider the timescale used in the 
N-body simulations. Although after 2.24 Gyr we see clear
morphological distinctions between the different types of bars, we 
have not trained the CNN on any images beyond this timescale. 
Another consequence of using the final time-step image as 
input to the CNN is that bars that survived the merger could have 
been destroyed in post-merger interactions before the end of the 
simulation. Although we have not 
seen direct visual evidence for this, we cannot rule out bar 
destruction several hundreds of Myr after the merger completes. 
Thus the obtained $f_{\text{bar}}$ values in Table \ref{tab:mbf} 
may have been higher were the CNN to have focused on the timestep 
immediately following the merger rather than the final timestep 
after around 2.24 Gyr. In practice, this is near impossible to 
implement for it either requires an accurate prediction of the 
merger timescale in advance, or the manual inspection of each model 
(defeating the purpose of using neural networks). Were a CNN to be 
trained using images from all timesteps, it may be possible to 
examine the rates of bar formation as a function of time, however 
the timestep must be small enough to capture the immediate 
aftermath of the galaxy merger.

Despite these issues, CNNs are an 
incredibly powerful tool that have the potential to revolutionise 
the speed with which large-scale imaging surveys can be processed 
and analysed. There are many benefits of deep learning-based 
approaches to observational studies, particularly since deep 
learning is a general paradigm whereby the neural networks can be 
easily tailored to the task at hand. Furthermore, a CNN that can 
classify barred galaxies according to the mechanism with which they 
formed would be a boon for studies into the environmental 
dependence of bars. Automated galaxy classification based on 
machine learning is a promising method with which to quickly parse 
large-scale data and accelerate studies into galaxy formation and 
evolution. However, as with all applications of machine learning, 
there is the important caveat that the
overall performance of the neural network is dependent on 
the accuracy, quality and relative size of the data on which it is 
trained \citep{haykins-2009}. As modern digital surveys yield 
increasingly larger datasets, the
suitability and accuracy of neural networks and other deep learning 
methods will only increase.

\subsection{Future considerations}

We established our results using N-body simulations rather than using hydrodynamic simulations that 
include gas. A key reason for this was to simplify the overall 
simulation, lowering the time it takes to evolve a single galaxy. 
This is important since, in order to obtain the trends in Figures 
\ref{fig:m-all} and \ref{fig:mfixed}, it was 
necessary to run thousands of models, with each N-body simulation 
taking a considerable amount of time to complete depending on 
stellar mass and bulge mass.
Adding gas would greatly increase the time taken to simulate 
each sample, making it impossible to carry out this work. It is 
known that gas plays an important role in the overall evolution of 
stellar bars \citep{berentzen-1998, bournard-2002, bournard-2005, 
berentzen-2007}, particularly in gas-rich isolated disks where the 
gas can act as a stabilising force, reduce the rate bar formation, 
and can result in weaker bar \citep{athana-2013}. However, since 
the main focus of this work is on bar formation in galaxy mergers 
where much gas stripping occurs, we have decided to use N-body 
simulations and omit tracking gas particles entirely.
Other future considerations include utilising and/or training a new  
CNN to be used with observational surveys in order to determine the 
prevalence of different types of bars throughout the Universe.  
Such studies could also determine how the bar fraction changes due 
to environment or redshift.

\section{Conclusions}

Through our N-body simulations, we have investigated the formation 
of bars via three different formation mechanisms; the isolated 
(spontaneous) model, the tidal interaction model, and the galaxy 
merger model. We have designed and trained a convolutional neural 
network to identify the presence of bars based on two-dimensional 
normalised density maps. We have investigated how the fraction of 
bars detected by our CNN, $f_{\text{bar}}$, changes as we vary the 
mass ratio $m_2$ and spin angles $\theta_1, \theta_2$. We have been 
able to determine the parameter ranges most conducive to bar 
formation for multiple models that differ in both 
dark matter content and bulge-to-disk ratios. Through 
extending our CNN to include multiple labels, we have shown that it 
is feasible for a CNN to classify a bar according to the mechanism 
with which it was formed.
There are several important 
conclusions that we reached based on our work:

\begin{enumerate}[(1)]

\item We have found that minor mergers (i.e those with $m_2 \approx 0.1$) with similar 
orientations ($\theta_1 \approx \theta_2$) are most conducive to 
bar formation.

\item Through visualising the bar formation process, we have found 
that there are two distinct phases.  The first phase involves the 
formation of a tidally induced bar due to pre-merger interactions 
as the two galaxies spiral into each other.  This is more likely to 
occur in the minor merger case as the merger timescale is longer. 
The second phase involves whether or not the bar survives the 
merger or, in the case of a destructive major merger, whether the 
bar can regenerate.  Minor merging typically preserves the bar, 
whilst major merging destroys the bar.  By showing that is is 
possible for a bar to regenerate in the case of equal spin angles $
\theta_1 = \theta_2 = 0$ degrees (Figure \ref{fig:majmer-eqth}), we 
infer that the transfer of angular momentum is key 
to the regeneration of the bar.

\item It is possible for strong bars to 
form in minor merging in cases where bars are otherwise suppressed 
in isolation due to the presence of a large stellar bulge 
(Figure \ref{fig:bulge-iso-mer}). This suggests that bulges 
may help to facilitate bar formation in galaxy merging.

\item We found that some galaxy models in Table \ref{tab:models} 
that failed to produce strong bars in the isolated case produced 
strong bars in galaxy merging. Thus galaxy merging can enhance bar 
formation. This has several implications, for instance bars at high 
redshift $z$ can be triggered by early minor merging. This is also 
important for SB0 galaxies where it is thought that merging is key 
to their formation.

\item We have shown that is feasible for a CNN to classify multiple 
bar formation mechanisms, however this 
was achieved using a much smaller training set than that employed 
for our main analysis. We found a higher accuracy when 
classifying isolated bars, albeit tidal and merger bars were more 
difficult to distinguish.

\end{enumerate}

\section*{Acknowledgements}

We are grateful to the referee for their comments and constructive feedback that helped to improve this paper.

\bibliographystyle{aa}
\bibliography{mybib} % if your bibtex file is called example.bib

\begin{thebibliography}{92}
\expandafter\ifx\csname natexlab\endcsname\relax\def\natexlab#1{#1}\fi

\bibitem[{Abraham {et~al.}(2018)Abraham, Aniyan, Kembhavi, Philip, \&
  Vaghmere}]{abraham-2018}
Abraham, S., Aniyan, A.~K., Kembhavi, A.~K., Philip, N.~S., \& Vaghmere, K.
  2018, MNRAS, 477, 894

\bibitem[{Aguerri {et~al.}(2009)Aguerri, M\'{e}ndez-Abreu, \&
  Corsini}]{aguerri-2009}
Aguerri, J., M\'{e}ndez-Abreu, J., \& Corsini, E. 2009, A\&A, 495, 491

\bibitem[{{Aguerri} {et~al.}(1998){Aguerri}, {Beckman}, \&
  {Prieto}}]{aguerri-1998}
{Aguerri}, J.~A.~L., {Beckman}, J.~E., \& {Prieto}, M. 1998, AJ, 116, 5

\bibitem[{Aguerri {et~al.}(2015)Aguerri, M{\'{e}}ndez-Abreu,
  Falc{\'{o}}n-Barroso, Amorin, Barrera-Ballesteros, Fernandes,
  Garc{\'{\i}}a-Benito, Garc{\'{\i}}a-Lorenzo, Delgado, Husemann, Kalinova,
  Lyubenova, Marino, M{\'{a}}rquez, Mast, P{\'{e}}rez, S{\'{a}}nchez, van~de
  Ven, Walcher, Backsmann, Cortijo-Ferrero, Bland-Hawthorn, del Olmo,
  Iglesias-P{\'{a}}ramo, P{\'{e}}rez, S{\'{a}}nchez-Bl{\'{a}}zquez, Wisotzki,
  \& Ziegler}]{aguerri-2015}
Aguerri, J. A.~L., M{\'{e}}ndez-Abreu, J., Falc{\'{o}}n-Barroso, J., {et~al.}
  2015, A\&A, 576, A102

\bibitem[{{Athanassoula}(1999)}]{athana-1999}
{Athanassoula}, E. 1999, in Astronomical Society of the Pacific Conference
  Series, Vol. 160, Astrophysical Discs - an EC Summer School, ed. J.~A.
  {Sellwood} \& J.~{Goodman}, 351

\bibitem[{Athanassoula(2003)}]{athana-2003}
Athanassoula, E. 2003, MNRAS, 341, 1179

\bibitem[{Athanassoula(2005)}]{athana-2005}
Athanassoula, E. 2005, Celestial Mech Dyn Astr, 91, 9

\bibitem[{{Athanassoula} {et~al.}(1983){Athanassoula}, {Bienayme}, {Martinet},
  \& {Pfenniger}}]{athana-1983}
{Athanassoula}, E., {Bienayme}, O., {Martinet}, L., \& {Pfenniger}, D. 1983,
  A\&A, 127, 349

\bibitem[{Athanassoula {et~al.}(2013)Athanassoula, Machado, \&
  Rodionov}]{athana-2013}
Athanassoula, E., Machado, R. E.~G., \& Rodionov, S.~A. 2013, MNRAS, 429, 1949

\bibitem[{Barber{\`{a}} {et~al.}(2004)Barber{\`{a}}, Athanassoula, \&
  Garc{\'{\i}}a-G{\'{o}}mez}]{barbera-2004}
Barber{\`{a}}, C., Athanassoula, E., \& Garc{\'{\i}}a-G{\'{o}}mez, C. 2004,
  A\&A, 415, 849

\bibitem[{{Barnes}(1996)}]{barnes-1996}
{Barnes}, J. 1996, in IAU Symposium, Vol. 171, New Light on Galaxy Evolution,
  ed. R.~{Bender} \& R.~L. {Davies}, 191

\bibitem[{Barnes \& Hernquist(1992)}]{barnes-1992}
Barnes, J.~E. \& Hernquist, L. 1992, ARA\&A, 30, 705

\bibitem[{Bekki(2013)}]{bekki-2013}
Bekki, K. 2013, MNRAS, 432, 2298

\bibitem[{Bekki {et~al.}(2019)Bekki, Diaz, \& Stanley}]{bekki-2019}
Bekki, K., Diaz, J., \& Stanley, N. 2019, A\&C, 2800286

\bibitem[{Berentzen {et~al.}(2004)Berentzen, Athanassoula, Heller, \&
  Fricke}]{berentzen-2004}
Berentzen, I., Athanassoula, E., Heller, C.~H., \& Fricke, K.~J. 2004, MNRAS,
  347, 220

\bibitem[{Berentzen {et~al.}(1998)Berentzen, Heller, Shlosman, \&
  Fricke}]{berentzen-1998}
Berentzen, I., Heller, C.~H., Shlosman, I., \& Fricke, K.~J. 1998, MNRAS, 300,
  49

\bibitem[{Berentzen {et~al.}(2007)Berentzen, Shlosman, Martinez-Valpuesta, \&
  Heller}]{berentzen-2007}
Berentzen, I., Shlosman, I., Martinez-Valpuesta, I., \& Heller, C.~H. 2007,
  ApJ, 666, 189

\bibitem[{Binney \& Tremaine(2008)}]{binney-2008}
Binney, J. \& Tremaine, S. 2008, Galactic Dynamics, 2nd edn. (Princeton
  University Press)

\bibitem[{Bournaud \& Combes(2002)}]{bournard-2002}
Bournaud, F. \& Combes, F. 2002, A\&A, 392, 83

\bibitem[{Bournaud {et~al.}(2005)Bournaud, Combes, \& Semelin}]{bournard-2005}
Bournaud, F., Combes, F., \& Semelin, B. 2005, MNRASL, 364, L18

\bibitem[{Bridge {et~al.}(2007)Bridge, Appleton, Conselice, Choi, Armus, Fadda,
  Laine, Marleau, Carlberg, Helou, \& Yan}]{bridge-2007}
Bridge, C.~R., Appleton, P.~N., Conselice, C.~J., {et~al.} 2007, ApJ, 659, 931

\bibitem[{Calleja \& Fuentes(2004)}]{calleja-2004}
Calleja, J. D.~L. \& Fuentes, O. 2004, MNRAS, 349, 87

\bibitem[{Cattaneo {et~al.}(2011)Cattaneo, Mamon, Warnick, \&
  Knebe}]{cattaneo-2011}
Cattaneo, A., Mamon, G.~A., Warnick, K., \& Knebe, A. 2011, A\&A, 533, A5

\bibitem[{Chollet {et~al.}(2015)}]{chollet-2015}
Chollet, F. {et~al.} 2015, Keras, \url{https://keras.io}

\bibitem[{Conselice(2014)}]{conselice-2014}
Conselice, C.~J. 2014, ARA\&A, 52, 291

\bibitem[{Consolandi(2016)}]{consolandi-2016}
Consolandi, G. 2016, A\&A, 595, A67

\bibitem[{Contopoulos \& Grosbøl(1989)}]{contopoulos-1989}
Contopoulos, G. \& Grosbøl, P. 1989, A\&AR, 1, 261

\bibitem[{Dalcanton {et~al.}(2004)Dalcanton, Yoachim, \&
  Bernstein}]{dalcanton-2004}
Dalcanton, J., Yoachim, P., \& Bernstein, R. 2004, ApJ, 608, 189

\bibitem[{Darg {et~al.}(2010)Darg, Kaviraj, Lintott, Schawinski, Sarzi,
  Bamford, Silk, Andreescu, Murray, Nichol, Raddick, Slosar, Szalay, Thomas, \&
  Vandenberg}]{darg-2010}
Darg, D.~W., Kaviraj, S., Lintott, C.~J., {et~al.} 2010, MNRAS, 401, 1552

\bibitem[{{Di Matteo} {et~al.}(2010){Di Matteo}, {Qu}, {Lehnert}, {van Driel},
  \& {Jog}}]{di-matteo-2010}
{Di Matteo}, P., {Qu}, Y., {Lehnert}, M.~D., {van Driel}, W., \& {Jog}, C.~J.
  2010, EAS Publications Series, 45, 389

\bibitem[{Diaz {et~al.}(2019)Diaz, Bekki, Forbes, Couch, Drinkwater, \&
  Deeley}]{diaz-2019}
Diaz, J.~D., Bekki, K., Forbes, D.~A., {et~al.} 2019, MNRAS, 486, 4845

\bibitem[{Dieleman {et~al.}(2015)Dieleman, Willett, \& Dambre}]{dieleman-2015}
Dieleman, S., Willett, K.~W., \& Dambre, J. 2015, MNRAS, 450, 1441

\bibitem[{Elmegreen {et~al.}(1990)Elmegreen, Bellin, \&
  Elmegreen}]{elmegreen-1990}
Elmegreen, D.~M., Bellin, A.~D., \& Elmegreen, B.~G. 1990, ApJ, 364, 415

\bibitem[{Eskridge \& Frogel(1999)}]{eskridge-1999}
Eskridge, P.~B. \& Frogel, J.~A. 1999, Ap\&SS, 269, 427

\bibitem[{Eskridge {et~al.}(2000)Eskridge, Frogel, Pogge, Quillen, Davies,
  Depoy, Houdashelt, Kuchinski, Ramírez, Sellgren, Terndrup, , \&
  Tiede}]{eskridge-2000}
Eskridge, P.~B., Frogel, J.~A., Pogge, R.~W., {et~al.} 2000, AJ, 119, 536

\bibitem[{Fanali {et~al.}(2015)Fanali, Dotti, Fiacconi, \&
  Haardt}]{fanali-2015}
Fanali, R., Dotti, M., Fiacconi, D., \& Haardt, F. 2015, MNRAS, 454, 3641

\bibitem[{Friedli \& Benz(1993)}]{friedli-1993}
Friedli, D. \& Benz, W. 1993, A\&A, 268, 65

\bibitem[{Gadotti(2011)}]{gadotti-2011}
Gadotti, D. 2011, MNRAS, 415, 3308

\bibitem[{{Garcia-G\'{o}mez} \& {Athanassoula}(1991)}]{garcia-gomez-1991}
{Garcia-G\'{o}mez}, C. \& {Athanassoula}, E. 1991, A\&AS, 89, 159

\bibitem[{Garcia-G{\'{o}}mez {et~al.}(2017)Garcia-G{\'{o}}mez, Athanassoula,
  Barber{\`{a}}, \& Bosma}]{garcia-gomez-2017}
Garcia-G{\'{o}}mez, C., Athanassoula, E., Barber{\`{a}}, C., \& Bosma, A. 2017,
  A\&A, 601, 132

\bibitem[{Gerssen {et~al.}(2003)Gerssen, Kuijken, \& Merrifield}]{gerssen-2003}
Gerssen, J., Kuijken, K., \& Merrifield, M.~R. 2003, MNRAS, 345, 261

\bibitem[{Goldreich \& Lynden-Bell(1965)}]{goldreich-1965}
Goldreich, P. \& Lynden-Bell, D. 1965, MNRAS, 130, 125

\bibitem[{Graff {et~al.}(2014)Graff, Feroz, Hobson, \& Lasenby}]{graff-2014}
Graff, P., Feroz, F., Hobson, M.~P., \& Lasenby, A. 2014, MNRAS, 441, 1741

\bibitem[{Haykin(2009)}]{haykins-2009}
Haykin, S.~O. 2009, Neural Networks and Learning Machines, 3rd edn. (Pearson)

\bibitem[{{Hernquist} \& {Mihos}(1995)}]{hernquist-1995}
{Hernquist}, L. \& {Mihos}, J.~C. 1995, ApJ, 448, 41

\bibitem[{Hohl(1971)}]{hohl-1971}
Hohl, F. 1971, ApJ, 168, 343

\bibitem[{Huertas-Company {et~al.}(2019)Huertas-Company, Rodriguez-Gomez,
  Nelson, Pillepich, Bottrell, Bernardi, Dom{\'{\i}}nguez-S{\'{a}}nchez, Genel,
  Pakmor, Snyder, \& Vogelsberger}]{huertascompany-2019}
Huertas-Company, M., Rodriguez-Gomez, V., Nelson, D., {et~al.} 2019, MNRAS,
  489, 1859

\bibitem[{Julian \& Toomre(1966)}]{julian-1966}
Julian, W.~H. \& Toomre, A. 1966, ApJ, 146, 810

\bibitem[{Kataria \& Das(2017)}]{kataria-2017}
Kataria, S.~K. \& Das, M. 2017, MNRAS, 475, 1653

\bibitem[{Kaviraj {et~al.}(2015)Kaviraj, Devriendt, Dubois, Slyz, Welker,
  Pichon, Peirani, \& Borgne}]{kaviraj-2015}
Kaviraj, S., Devriendt, J., Dubois, Y., {et~al.} 2015, MNRAS, 452, 2845

\bibitem[{Kaviraj {et~al.}(2009)Kaviraj, Peirani, Khochfar, Silk, \&
  Kay}]{kaviraj-2009}
Kaviraj, S., Peirani, S., Khochfar, S., Silk, J., \& Kay, S. 2009, MNRAS, 394,
  1713

\bibitem[{Kormendy \& Kennicutt(2004)}]{kormendy-2004}
Kormendy, J. \& Kennicutt, Jr., R.~C. 2004, AR\&A, 42, 603

\bibitem[{LeCun {et~al.}(2015)LeCun, Bengio, \& Hinton}]{lecun-2015}
LeCun, Y., Bengio, Y., \& Hinton, G. 2015, Nature, 512, 436

\bibitem[{LeCun {et~al.}(1998)LeCun, Bottou, Bengio, \& Haffner}]{lecun-1998}
LeCun, Y., Bottou, L., Bengio, Y., \& Haffner, P. 1998, Proceedings of the
  {IEEE}, 86, 2278

\bibitem[{Lin {et~al.}(2007)Lin, Koo, Weiner, Chiueh, Coil, Lotz, Conselice,
  Willner, Smith, Guhathakurta, Huang, Floc'h, Noeske, Willmer, Cooper, \&
  Phillips}]{lin-2007}
Lin, L., Koo, D.~C., Weiner, B.~J., {et~al.} 2007, ApJ, 660, L51

\bibitem[{Little \& Carlberg(1991)}]{little-1991}
Little, B. \& Carlberg, R.~G. 1991, MNRAS, 250, 161

\bibitem[{Lotz {et~al.}(2010)Lotz, Jonsson, Cox, \& Primack}]{lotz-2010}
Lotz, J.~M., Jonsson, P., Cox, T.~J., \& Primack, J.~R. 2010, MNRAS, 404, 575

\bibitem[{Lukic {et~al.}(2019)Lukic, Br\"{u}ggen, Mingo, Croston, Kasieczka, \&
  Best}]{lukic-2019}
Lukic, V., Br\"{u}ggen, M., Mingo, B., {et~al.} 2019, MNRAS, 487, 1729

\bibitem[{Lynden-Bell(1979)}]{lynden-bell-1979}
Lynden-Bell, D. 1979, MNRAS, 187, 101

\bibitem[{Lynden-Bell \& Kalnajs(1972)}]{lynden-bell-1972}
Lynden-Bell, D. \& Kalnajs, A.~J. 1972, MNRAS, 157, 1

\bibitem[{Mayer {et~al.}(2006)Mayer, Mastropietro, Wadsley, Stadel, \&
  Moore}]{mayer-2006}
Mayer, L., Mastropietro, C., Wadsley, J., Stadel, J., \& Moore, B. 2006, MNRAS,
  369, 1021

\bibitem[{Miwa \& Noguchi(1998)}]{miwa-1998}
Miwa, T. \& Noguchi, M. 1998, ApJ, 499, 1

\bibitem[{{Moetazedian} {et~al.}(2017){Moetazedian}, {Polyachenko}, {Berczik},
  \& {Just}}]{moetazedian-2017}
{Moetazedian}, R., {Polyachenko}, E.~V., {Berczik}, P., \& {Just}, A. 2017,
  A\&A, 604, A75

\bibitem[{Naim {et~al.}(1995)Naim, Lahav, Sodre, \&
  Storrie-Lombardi}]{naim-1995}
Naim, A., Lahav, O., Sodre, L., \& Storrie-Lombardi, M.~C. 1995, MNRAS, 275,
  567

\bibitem[{Navarro {et~al.}(1996)Navarro, Frenk, \& White}]{nfw}
Navarro, J.~F., Frenk, C.~S., \& White, S. D.~M. 1996, ApJ, 462, 563

\bibitem[{{Negroponte} \& {White}(1983)}]{negroponte-1983}
{Negroponte}, J. \& {White}, S.~D.~M. 1983, MNRAS, 205, 1009

\bibitem[{Neto {et~al.}(2007)Neto, Gao, Bett, Cole, Navarro, Frenk, White,
  Springel, \& Jenkins}]{neto-2007}
Neto, A.~F., Gao, L., Bett, P., {et~al.} 2007, MNRAS, 381, 1450

\bibitem[{Noguchi(1987)}]{noguchi-1987}
Noguchi, M. 1987, MNRAS, 228, 635

\bibitem[{Pearson {et~al.}(2019)Pearson, Wang, Alpaslan, Baldry, Bilicki,
  Brown, Grootes, Holwerda, Kitching, Kruk, \& van~der Tak}]{pearson-2019}
Pearson, W.~J., Wang, L., Alpaslan, M., {et~al.} 2019, A\&A, 631, A51

\bibitem[{Pedrosa \& Tissera(2015)}]{pedrosa-2015}
Pedrosa, S.~E. \& Tissera, P.~B. 2015, A\&A, 584, A43

\bibitem[{Peirani {et~al.}(2009)Peirani, Hammer, Flores, Yang, \&
  Athanassoula}]{peirani-2009}
Peirani, S., Hammer, F., Flores, H., Yang, Y., \& Athanassoula, E. 2009, A\&A,
  496, 51

\bibitem[{P{\'{e}}rez {et~al.}(2017)P{\'{e}}rez, Mart{\'{\i}}nez-Valpuesta,
  Ruiz-Lara, de~Lorenzo-Caceres, Falc{\'{o}}n-Barroso, Florido, Delgado,
  Lyubenova, Marino, S{\'{a}}nchez, S{\'{a}}nchez-Bl{\'{a}}zquez, van~de Ven,
  \& Zurita}]{perez-2017}
P{\'{e}}rez, I., Mart{\'{\i}}nez-Valpuesta, I., Ruiz-Lara, T., {et~al.} 2017,
  MNRAS:Letters, 470, L122

\bibitem[{Pfenniger(1984)}]{pfenniger-1984}
Pfenniger, D. 1984, A\&A, 134, 373

\bibitem[{Pfenniger \& Friedli(1991)}]{pfenniger-1991}
Pfenniger, D. \& Friedli, D. 1991, A\&A, 252, 75

\bibitem[{Prieto {et~al.}(2001)Prieto, Aguerri, Varela, \& {n}oz
  Tu\~{n}\'{o}n}]{prieto-2001}
Prieto, M., Aguerri, J., Varela, A., \& {n}oz Tu\~{n}\'{o}n, C.~M. 2001, A\&A,
  367, 405

\bibitem[{Raha {et~al.}(1991)Raha, Sellwood, James, \& Kahn}]{raha-1991}
Raha, N., Sellwood, J.~A., James, R.~A., \& Kahn, F.~D. 1991, Nature, 352, 411

\bibitem[{Rautiainen {et~al.}(2002)Rautiainen, Salo, \&
  Laurikainen}]{rautiainen-2002}
Rautiainen, P., Salo, H., \& Laurikainen, E. 2002, MNRAS, 337, 1233

\bibitem[{Regan \& Teuben(2003)}]{regan-2003}
Regan, M.~W. \& Teuben, P. 2003, ApJ, 582, 723

\bibitem[{Reichard {et~al.}(2008)Reichard, Heckman, Rudnick, Brinchmann, \&
  Kauffmann}]{reichard-2008}
Reichard, T.~A., Heckman, T.~M., Rudnick, G., Brinchmann, J., \& Kauffmann, G.
  2008, ApJ, 677, 186

\bibitem[{Sellwood(2014)}]{sellwood-2014}
Sellwood, J.~A. 2014, Rev. Mod. Phys., 86, 1

\bibitem[{Sellwood \& Sparke(1988)}]{sellwood-1998}
Sellwood, J.~A. \& Sparke, L.~S. 1988, MNRAS, 231, 25P

\bibitem[{Sellwood \& Wilkinson(1993)}]{sellwood-1993}
Sellwood, J.~A. \& Wilkinson, A. 1993, Rep. Prog. Phys, 56, 173

\bibitem[{Sparke \& Sellwood(1987)}]{sparke-1987}
Sparke, L.~S. \& Sellwood, J.~A. 1987, MNRAS, 225, 653

\bibitem[{Springel \& Hernquist(2005)}]{springel-2005}
Springel, V. \& Hernquist, L. 2005, ApJL, 622, L9

\bibitem[{Toomre(1964)}]{toomre-1964}
Toomre, A. 1964, ApJ, 139, 1217

\bibitem[{Toomre \& Toomre(1972)}]{toomre-1972}
Toomre, A. \& Toomre, J. 1972, ApJ, 178, 623

\bibitem[{Tremaine \& Weinberg(1984)}]{tremaine-1984}
Tremaine, S. \& Weinberg, M.~D. 1984, ApJ, 282, L5

\bibitem[{Vasconcellos {et~al.}(2011)Vasconcellos, de~Carvalho, Gal, LaBarbera,
  Capelato, Velho, Trevisan, \& Ruiz}]{vasconcellos-2011}
Vasconcellos, E.~C., de~Carvalho, R.~R., Gal, R.~R., {et~al.} 2011, AJ, 141,
  189

\bibitem[{Vera {et~al.}(2016)Vera, Alonso, \& Coldwell}]{vera-2016}
Vera, M., Alonso, S., \& Coldwell, G. 2016, A\&A, 595, A63

\bibitem[{Wu {et~al.}(2018{\natexlab{a}})Wu, Wong, Rudnick, Shabala, Alger,
  Banfield, Ong, White, Garon, Norris, Andernach, Tate, Lukic, Tang,
  Schawinski, \& Diakogiannis}]{wu-2018}
Wu, C., Wong, O.~I., Rudnick, L., {et~al.} 2018{\natexlab{a}}, MNRAS, 482, 1211

\bibitem[{Wu {et~al.}(2018{\natexlab{b}})Wu, Pfenniger, \&
  Taam}]{wuyuting-2018}
Wu, Y.-T., Pfenniger, D., \& Taam, R.~E. 2018{\natexlab{b}}, ApJ, 860, 152

\bibitem[{Zeiler(2012)}]{zeiler-2012}
Zeiler, M.~D. 2012, astro-ph/arXiv:1212.5701

\end{thebibliography}

\end{document}